\documentclass{article}
\usepackage{slashed,verbatim,subfigure}
\usepackage[numbers,sort&compress]{natbib}
\usepackage{amsmath}
\usepackage{amssymb}
\usepackage{appendix}
\usepackage{graphicx}
\usepackage{dcolumn}
\usepackage{bm}
\usepackage[colorlinks=true,linktocpage=true,
linkcolor=blue,citecolor=blue]{hyperref}
\usepackage[a4paper]{geometry}
\newcommand{\be}{\begin{eqnarray}}
\newcommand{\ee}{\end{eqnarray}}
\newcommand{\dex}{f^{\rm q,\rm ex}_{\rm aniso}}
\newcommand{\db}{f^{\rm q,\rm B}_{\rm aniso}}
\newcommand{\df}{f^{\rm q}_{\rm iso}}
\newcommand{\dg}{f^{\rm g}_{\rm iso}}
\newcommand{\dgex}{f^{\rm g,\rm ex}_{\rm aniso}}
\begin{document}
\large
\title{\bf{Momentum and its affiliated transport coefficients 
for a hot QCD matter in a strong magnetic field}}
\author{Shubhalaxmi Rath\footnote{srath@ph.iitr.ac.in}~~and~~Binoy Krishna 
Patra\footnote{binoy@ph.iitr.ac.in}\vspace{0.03in} \\ 
Department of Physics, Indian Institute of Technology Roorkee, Roorkee 247667, India}
\date{}
\maketitle
\begin{abstract}
We have studied the effects of anisotropies on the momentum transport in
a strongly interacting matter by the transport coefficients, {\em viz.} 
shear ($\eta$) and bulk ($\zeta$) viscosities. The anisotropies could 
arise either by the strong magnetic field or by the preferential 
expansion, both of which are created in the very early stages of 
ultrarelativistic heavy 
ion collisions at RHIC or LHC. This study is thereby aimed to 
understand (i) the fluidity and location of transition point 
of the matter through $\eta/s$ and $\zeta/s$ ($s$ is the entropy density), 
respectively, (ii) the sound attenuation through the Prandtl number (Pl), 
(iii) the nature of the flow by the Reynolds number (Rl), and (iv) the 
competition between momentum and charge diffusions through the 
ratio $({\eta}/{s})/({\sigma_{\rm el}}/{T})$. For this purpose, 
we have first calculated the viscosities in the relaxation-time 
approximation of kinetic theory approach and the interactions among 
partons are embodied by assigning masses to quarks and gluons
at finite temperature and strong magnetic field, known as quasiparticle 
model. Compared to the isotropic medium, both $\eta$ and $\zeta$ get 
increased in the magnetic field-driven ($B$-driven) anisotropy, contrary to the 
decrease in the expansion-driven anisotropy. Zooming in, $\eta$ increases 
with temperature faster in the former case than in the latter case, whereas 
$\zeta$ in the former case monotonically 
decreases with the temperature and in the latter case, it is meagre 
and ultimately diminishes at a specific temperature. Thus, the 
behaviors of shear and bulk viscosities could in principle distinguish the 
aforesaid anisotropies. As a result, $\eta/s$ gets enhanced in the former 
case but decreases with temperature and in the latter case, it becomes 
even smaller than the isotropic one. Similarly, $\zeta/s$ gets amplified 
but decreases faster with the temperature in the presence of strong 
magnetic field. The Prandtl number gets 
increased in $B$-induced anisotropy and gets decreased in 
expansion-induced anisotropy, compared to the isotropic case. However, Pl 
is always found larger than 1, so the sound attenuation is mostly 
governed by the momentum diffusion. The momentum anisotropy due to the 
magnetic field makes the Reynolds number smaller than 1, 
whereas the expansion-driven anisotropy makes it larger. Finally the 
ratio $({\eta}/{s})/({\sigma_{\rm el}}/{T})$ is amplified much 
in the presence of magnetic field-driven anisotropy, whereas the 
amplification is less pronounced in isotropic medium as well as in 
expansion-driven anisotropic medium. However, the ratio is always more than 1, 
so the momentum diffusion always prevails over the charge diffusion. 

\end{abstract}


\newpage

\section{Introduction}
Ultrarelativistic heavy-ion collisions (URHICs) at RHIC and LHC provide an 
enticing opportunity to investigate the strongly interacting matter in the 
form of deconfined quarks and gluons, dubbed as quark-gluon plasma (QGP). 
One of the amazing findings at RHIC and LHC is the substantial 
collective flow and the data are well-reproduced by perfect fluid  dynamics~\cite{Appelshauser:PRL80'1998}. In a parallel theoretical discovery, a 
lower bound ($1/4\pi$) in the ratio of shear viscosity ($\eta$) to 
entropy density ($s$) is found for some physical systems, {\em such as}
quarks and gluons, helium, nitrogen and water at and near their 
phase transitions~\cite{Kovtun:PRL94'2005,Shuryak:NPA750'2005}. 
Conversely, there are indications that the ratio of bulk viscosity ($\zeta$) to 
entropy density may have a maximum in the vicinity of the phase transition. Thus, the location of the transition or rapid crossover in QCD via the 
ratios $\eta/s$ and $\zeta/s$ can be pinpointed, in addition to and 
independent of the equation of state. The abovementioned predictions 
were made for the simplest possible phenomenological setting, {\em i.e.} 
fully central collisions. However, an intensely strong magnetic field
is expected to be produced at very early stages of URHICs, when the events are
off-central~\cite{Kharzeev:NPA803'2008}. 
Depending on the centrality, the strength of the magnetic
field may reach between $m_\pi^2$ (${10}^{18}$ Gauss) at RHIC
to 15 $m_\pi^2$ at LHC~\cite{Skokov:IJMPA24'2009} and at extreme cases it may 
reach 50 $m_\pi^2$. Naive classical estimates predict that the magnetic 
field may be very strong for very short duration~\cite{McLerran:NPA929'2014}. 
However, the realistic calculations on the charge transport properties of 
the produced medium, {\em mainly} the electrical conductivity suggest that the 
magnetic field may remain substantially strong for significantly longer time \cite{Tuchin:PRC82'2010,Conductivities}. Since the abovementioned collective 
flow has been interpreted as strong indicator of early thermalization, the strong magnetic field created at the early stages of URHICs might affect the 
momentum transport of the produced matter.

A wide range of theoretical and phenomenological observations have been made 
on how the strong magnetic field influences the properties of hot QCD matter, 
{\em such as} thermodynamic and magnetic properties 
\cite{Thermodynamics,Rath:JHEP1712'2017,Bjorken expansion,Karmakar:PRD99'2019}, 
chiral magnetic effect \cite{Fukushima:PRD78'2008,Kharzeev:NPA803'2008},
dilepton production~\cite{Tuchin:PRC88'2013,Mamo:JHEP1308'2013},
(inverse) magnetic catalysis due to the (restoration) breaking of the chiral 
symmetry \cite{Mueller:PRD91'2015,Haber:PRD90'2014,Gusynin:PRL73'1994} etc. 
As an artifact of the strong magnetic field, the dynamics of quarks along the 
longitudinal direction ($p_L$) dominates over the motion along the 
transverse direction ($p_T$) ($p_L\gg p_T$). This is further evidenced 
in the quantum-mechanical dispersion relation for a flavor ($i$) 
of mass, $m_i$ and electric charge, 
$q_i$: $\omega_{i,n}=\sqrt{p_L^2+2n |q_iB|+m_i^2}$, 
where only the lowest Landau level ($n=0$) is populated in 
the strong magnetic field limit ($|q_iB| \gg T^2$ as well as
$|q_iB| \gg m_i^2$, abbreviated as SMF limit), {\em i.e.} the case of 
vanishingly small $p_T$ ($\approx 0$), thus it results in an anisotropy in the 
momentum space. For weak-momentum anisotropic limit 
($\xi=\frac{\langle {\bf p}_{T}^{2}\rangle}{2\langle 
p_{L}^{2}\rangle}-1 <1$), the anisotropic distribution function for 
quark could be conceived by stretching the isotropic distribution 
in the direction of anisotropy.\footnote{It is worth to mention 
here that, although gluons being uncharged particles are not 
directly affected by the magnetic field-driven anisotropy, but their 
dynamics can be indirectly influenced by the magnetic field through the 
modification of the Debye screening mass.} Another kind of momentum 
anisotropy could also emerge at the similar time scale of magnetic field 
production due to the asymptotic free expansion of the matter along the beam 
direction compared to its transverse direction ($p_T\gg p_L$)
~\cite{Romatschke:PRD68'2003}. So, unlike the aforesaid anisotropy, the 
(weak) anisotropic distribution functions for both quarks and gluons 
could be approximated by contracting the respective distribution 
functions in the direction of anisotropy due to the positive value of $\xi$. 

In order to take into account the dissipative processes, {\em namely} 
thermal conduction and viscosity etc., one usually goes to the next 
approximation beyond the initial local equilibrium distribution function 
($f_0$), {\em i.e.} $f=f_0+\delta f$. The correction, 
$\delta f$ is determined by solving the transport equation, after linearizing 
the collision integral (which also involves the initial local 
equilibrium distribution function) with respect to the correction. Thus, if the 
initial distribution is anisotropic then the initial anisotropy is 
going to affect the solution of transport equation, which in turn 
affects the transport coefficients. If the medium exhibits weak anisotropy 
then the transport coefficients are decomposable into isotropic and 
anisotropic terms. For an example, due to the asymptotic expansion at very high energy 
in the early stages of the collisions, the expansion rate along the 
longitudinal direction becomes much higher than that along the transverse 
direction. As a result, the system becomes much colder in the 
longitudinal direction than in the transverse direction, 
which gets translated into an anisotropy in the particle momentum 
distribution \cite{Dumitru:PRD79'2009,Gale:IJMPA28'2013}. Thus, the 
anisotropy initially present in the spatial distribution is translated 
into the anisotropy in the momentum distribution of 
particles \cite{Ollitrault:PRD46'1992,Luzum:JPG41'2014,Busza:ARNPS68'2018}. 
It is seen in the hydrodynamics study \cite{Song:PRC81'2010}, how a spatial 
anisotropy gets converted into a flow anisotropy in the momentum space 
for an expanding matter with finite shear and bulk viscosities. 
One might thus expect that, the anisotropies discussed hereinabove 
could affect the transport properties of the medium. Recently we
had explored the effects of aforesaid momentum anisotropies on the transports 
of charge and heat by electrical ($\sigma_{\rm el}$) and thermal ($\kappa$) conductivities, respectively, where not only their magnitudes have undergone a 
drastic change, but their behaviors are also seen a marked difference in 
abovementioned anisotropies~\cite{Conductivities}. Moreover we had 
also studied the affiliated coefficients related to $\sigma_{\rm el}$ and $\kappa$ 
by the Lorenz number in Wiedemann-Franz law and the Knudsen number, whose 
magnitudes as well as behaviors distinguish the anisotropies. As a 
corollary, the electrical conductivity thus obtained enhances the duration 
for which the magnetic field remains strong. 
In the present work, we intend to explore the effects of aforesaid 
anisotropies on the momentum transports across and along the layer 
by shear and bulk viscosities, respectively. This exploration will 
further facilitate to understand the effects of strong magnetic
field on the affiliated coefficients: (i) to check the fluidity 
and the transition point of the hot QCD matter by the ratios $\eta/s$ and 
$\zeta/s$, (ii) to observe the sound attenuation in the 
medium by the Prandtl number (Pl=$\frac{\eta C_p}{\rho \kappa}$, 
$C_p$: specific heat at constant pressure, $\rho$: mass density, 
$\kappa$: thermal conductivity), 
(iii) to characterize the nature of flow by the Reynolds number 
(Rl=$\frac{Lv\rho}{\eta}$, $L$ and $v$: characteristic length and velocity 
of the flow), and finally (iv) the competition between the momentum and charge 
diffusions by the ratio $(\eta/s)/(\sigma_{\rm el}/T)$. 
The studies on the abovementioned transport coefficients 
are helpful to understand the transport phenomena in other areas 
where strong magnetic fields might exist, {\em such as} the core 
of the magnetar and the beginning of the universe, in addition to URHICs.

A variety of calculations on shear and bulk viscosities have been 
done by applying the perturbation theory \cite{Arnold:JHEP'2000'2003,Arnold:PRD74'2006,Hidaka:PRD78'2008}, the 
kinetic theory \cite{Danielewicz:PRD31'1985,Sasaki:PRC79'2009,Thakur:PRD95'2017} 
etc. for a thermal medium of quarks and gluons in the absence of 
magnetic field. In the presence of magnetic field the rotational 
invariance is broken, which in turn induces an azimuthal 
anisotropy of produced particles. As a result, the viscous stress tensor 
is characterized by seven viscous coefficients, out of which five are 
shear viscosities and the remaining two are bulk viscosities 
\cite{Lifshitz:BOOK'1981,Huang:PRD81'2010,Tuchin:JPG39'2012,Critelli:PRD90'2014,
Hernandez:JHEP05'2017,Hattori:PRD96'2017,Chen:PRD101'2020}. Since the 
components of fluid velocity transverse to the magnetic field direction 
tend to zero~\cite{Tuchin:JPG39'2012}, {\em i.e.} they 
decay with a finite relaxation-time even in a zero spatial 
gradient limit, so, they are no longer long-lived hydrodynamic 
variables \cite{Li:PRD97'2018}. Specifically, in SMF limit, 
only the longitudinal components of shear and bulk viscosities 
along the direction of magnetic field survive, 
which are contributed only by the lowest Landau 
level (LLL) quarks/antiquarks, and other components become negligible  \cite{Lifshitz:BOOK'1981,Tuchin:JPG39'2012,Hattori:PRD96'2017}. The 
influence of magnetic field on the viscosities has also been 
investigated previously in various approaches and models, 
{\em such as}, correlator technique using Kubo formula 
\cite{Nam:PRD87'2013,Hattori:PRD96'2017}, perturbative 
QCD in weak magnetic field \cite{Li:PRD97'2018}, Chapman-Enskog 
method with effective fugacity approach 
\cite{Kurian:EPJC79'2019} and holographic model \cite{Rebhan:PRL108'2012,Jain:JHEP10'2015,Finazzo:PRD94'2016}. 
In our present work, we are going to calculate both the viscosities 
in both magnetic field- and expansion-driven anisotropies within the 
kinetic theory approach in the relaxation-time approximation. 
We will further examine the influence of anisotropies on the relative 
behavior between them by the abovementioned derived transport 
coefficients: $\eta/s$ and $\zeta/s$; the Prandtl number; the Reynolds number, 
and the ratio of momentum diffusion to charge diffusion, which are worthy 
of investigation for different perspectives. The ratio, $\eta/s$ is 
studied in holographic model by Kovtun, Son and 
Starinets~\cite{Kovtun:PRL94'2005} and reports a lower bound 
$\frac{1}{4\pi}$, irrespective of physical systems. The above ratio 
is also studied in parton transport model to reproduce the collective 
behavior \cite{Xu:PRL101'2008,Ferini:PLB670'2009,
Cassing:NPA831'2009, Bratkovskaya:NPA856'2011} at URHICs and is found to be 
very small ($\approx \frac{1}{4\pi}$) and hydrodynamic model 
\cite{Luzum:PRC78'2008} also reports the value of $\eta/s$ between 
$\frac{1}{4\pi}$ to $\frac{2}{4\pi}$ compatible with the 
experiments~\cite{Gavin:PRL97'2006,Drescher:PRC76'2007} as well as with 
the lattice calculations \cite{Nakamura:PRL94'2005,Meyer:PRD76'2007}. 
The ratio, $\zeta/s$ is found to be very small ($<0.15$) in lattice 
calculations~\cite{Meyer:PRL100'2008,Karsch:PLB663'2008} except 
for small region around the QCD deconfinement transition temperature 
$T_c$ and even becomes extremely small away from $T_c$. The Prandtl number 
is calculated for a strongly coupled liquid helium using 
kinetic theory \cite{T:RPP72'2009}, which is found to be around 2.5, and 
for a nonrelativistic conformal holographic fluid, Pl is 1.0 
\cite{T:RPP72'2009, Rangamani:JHEP01'2009}. 
For dilute atomic Fermi gas at high temperatures, Pl is 
calculated in the framework of kinetic theory, where it turns out to be 
$\frac{2}{3}$~\cite{Braby:PRA82'2010}. The magnitude of the Reynolds number 
indicates the type of flow, whether it is laminar (${\rm Rl} \leq 1$) 
or turbulent (${\rm Rl} \gg 1$)~\cite{McInnes:NPB921'2017}. The 
(3+1)-dimensional fluid dynamical model reports the value of the Rl for 
QGP in the range 3-10 \cite{Csernai:PRC85'2012}, whereas the holographic 
setup estimates the higher value as approximately 20 
\cite{McInnes:NPB921'2017}. Thus, the QGP is thought to be a viscous 
medium and the flow remains laminar. 
Similarly in the calculations using the relativistic kinetic theory  \cite{Thakur:PRD95'2017} and the Chapman-Enskog method with effective 
fugacity approach \cite{Mitra:PRD96'2017}, the ratio, 
$\gamma=(\eta/s)/(\sigma_{el}/T)$ is reported between 1 to 20 or 
even higher for QGP system near transition temperature and gets 
saturated at higher temperatures. 

Recently we have noticed that the noninteracting 
description of particles yields the unusually large values of thermal 
and electrical conductivities. So we have circumvented the problem by the 
quasiparticle description of particles, commonly known as quasiparticle 
model (QPM), where the interactions among the constituents are embodied 
in terms of the medium generated masses in the distribution functions of 
particles in the phase space. The QPM has been proposed previously in 
different approaches, {\em  such as} the Nambu-Jona-Lasinio (NJL) and 
Polyakov NJL-based quasiparticle models 
\cite{Fukushima:PLB591'2004,Ghosh:PRD73'2006,Abuki:PLB676'2009}, 
quasiparticle model with Gribov-Zwanziger quantization 
\cite{Su:PRL114'2015,Florkowski:PRC94'2016}, thermodynamically consistent 
quasiparticle model \cite{Bannur:JHEP0709'2007} etc. In this work, 
we have used the resummed propagators for quarks and gluons immersed in a thermal 
medium in the absence and in the presence of strong magnetic field by the 
respective self-energies and finally the poles of respective 
propagators yield the medium generated (quasiparticle) masses 
for quarks and gluons. With this quasiparticle description, the thermal 
and electrical conductivities were found finite~\cite{Conductivities}, but 
larger in the anisotropy induced by the strong magnetic field than by the 
expansion. Here also, in the magnetic field-driven ($B$-driven) anisotropy, not 
only the magnitude of shear viscosity becomes larger than that in the 
expansion-driven anisotropy, but its increase with temperature also 
becomes faster. Similarly the bulk viscosity is also larger in 
$B$-driven anisotropy but decreases slowly with the temperature, whereas 
in expansion-driven anisotropy, $\zeta$ is very small and abruptly 
approaches zero at a higher temperature ($>T_c$). 
Although the magnitude of the entropy density and its variation 
with the temperature get decreased in $B$-induced anisotropy compared to 
isotropic and expansion-induced anisotropic cases, but the increase of $\eta$ 
with temperature ($T$) is smaller than the increase of $s$ with $T$. As 
a result, unlike $\eta$ and $s$, $\eta/s$ decreases with temperature, 
but its magnitude is always larger than those in isotropic medium as 
well as in expansion-induced anisotropy. On the other hand, $\zeta/s$ 
gets enhanced in $B$-driven anisotropy, but it now decreases faster 
with temperature. The Prandtl number becomes 
higher in the $B$-driven anisotropy than that 
in the isotropic medium, whereas the expansion-driven anisotropy 
reduces this number to the value lower than that in the isotropic 
medium, thus showing opposite behavior in two anisotropies. 
However, in all cases the Prandtl number remains greater than 1, 
so the sound attenuation in an interacting system is mostly 
governed by the momentum diffusion. The Reynolds number 
becomes less than 1 in $B$-driven anisotropy, so the 
kinematic viscosity ($\eta/\rho$) dominates over the size and 
velocity of the flow and it describes the hot QCD matter as a 
viscous fluid, whereas in expansion-driven anisotropy Rl becomes 
greater than 1. Finally, we have observed that, the 
ratio $(\eta/s)/(\sigma_{el}/T)$ in $B$-driven anisotropy 
gets increased as compared to the isotropic one, but in the presence 
of expansion-driven anisotropy this ratio becomes smaller than the 
isotropic one. However, $(\eta/s)/(\sigma_{el}/T)$ is always 
larger than 1, therefore the momentum diffusion dominates over the 
charge diffusion. 

Our work is organized as follows. In section 2, we have reviewed the 
quasiparticle description of hot quarks and gluons in an ambience 
of strong magnetic field. Section 3 overall deals with the momentum transports 
by shear and bulk viscosities and their ratios with the entropy density. To 
be specific, in subsection 3.1, we have first revisited the shear and bulk 
viscosities in isotropic thermal medium and the same in the presence of the 
expansion- and strong magnetic field-induced anisotropies are computed in 
subsection 3.2. After computing the viscosities, we have 
calculated the ratios $\eta/s$ and $\zeta/s$ in subsection 3.3. In 
section 4, we have studied the coefficients affiliated to momentum, heat and 
charge transports through the Prandtl number, the Reynolds number and the ratio 
of momentum diffusion to charge diffusion. Finally, in section 5, we have concluded. 

\section{Quasiparticle description of partons at finite $T$ and strong $B$}
Quasiparticle description of quarks and gluons at finite temperature 
in the presence of magnetic field embodies the interactions 
among themselves in the form of thermal masses. Especially, different flavors 
acquire masses differently due to their different electric charges, in 
addition to their current masses. The masses are generated due to the 
interaction of a given parton in a given environment with other particles 
of the medium, therefore quasiparticle description in turn describes the 
collective properties of the medium. Different versions of quasiparticle 
description exist in the literature based on different effective theories, 
{\em such as} Nambu-Jona-Lasinio (NJL) model and its extension PNJL model~\cite{Fukushima:PLB591'2004,Ghosh:PRD73'2006,Abuki:PLB676'2009}, 
Gribov-Zwanziger quantization \cite{Su:PRL114'2015,Florkowski:PRC94'2016}, thermodynamically consistent quasiparticle model \cite{Bannur:JHEP0709'2007} 
etc. However, our description relies on perturbative thermal QCD, where the 
medium generated masses for quarks and gluons are obtained from the poles 
of dressed propagators calculated by the respective self-energies at 
finite temperature and/or strong magnetic field~\cite{Conductivities}.

Let us start with the quasiparticle description of quarks and gluons
in a thermal medium alone, where gluon acquires a thermal
mass~\cite{Bellac:BOOK'1996,Peshier:PRD66'2002},
\be\label{Gluon mass}
m_{gT}^2(T)=\frac{g^{\prime2}T^2}{6}\left(N_c+\frac{N_i}{2}\right)
\ee
and the quark also acquires a thermal mass,
\begin{eqnarray}\label{Quark mass}
m_{qT}^2 (T) =\frac{g^{\prime2}T^2}{6}
,\end{eqnarray}
where $g^\prime$ is the running coupling taken up to one-loop, which runs 
only with the temperature with the renormalization scale fixed at 
$2 \pi T$ and has the following \cite{BOOK} form,
\begin{eqnarray}
g^{\prime2}=\frac{48\pi^2}{\left(11N_c-2N_i\right)\ln\left({\Lambda^2}/{\Lambda_{\rm\overline{MS}}^2}\right)}
~,\end{eqnarray}
where $\Lambda=2 \pi T$ and $\Lambda_{\rm\overline{MS}}=0.176$ GeV. 

In the presence of strong magnetic field, the gluons are not affected 
directly by the magnetic field. However, the quark-loop of the gluon 
self-energy will be affected by the magnetic field, which in turn 
could affect the aforesaid mass \eqref{Gluon mass} 
\cite{Bjorken expansion,Fukushima:PRD93'2016,Singh:PRD97'2018} as 
\be\label{Gluon mass(eb)}
m_{gT,B}^2 (T,B)=\frac{g^{\prime2}T^2N_c}{6}+\frac{g^2}{8\pi^2}\sum_i|q_iB|
.\ee
We are now going to discuss the thermal quark mass in the presence 
of strong magnetic field, which will be given from the pole 
($p_0=0, \mathbf{p}\rightarrow 0$ limit) of the effective quark 
propagator. The effective propagator can be obtained self-consistently 
from the Schwinger-Dyson equation, which is given by
\be\label{S.D.E.}
S^{-1}(p_\parallel)=\gamma^\mu p_{\parallel\mu}-\Sigma(p_\parallel)
~,\ee
where $\Sigma (p_\parallel)$ is the quark self-energy at 
finite temperature in the presence of strong magnetic 
field. We can evaluate it up to one-loop from the 
following expression:
\begin{eqnarray}\label{Q.S.E.}
\Sigma(p)=-\frac{4}{3} g^{2}i\int{\frac{d^4k}{(2\pi)^4}}
\left[\gamma_\mu {S(k)}\gamma_\nu {D^{\mu \nu} (p-k)}\right]
,\end{eqnarray}
where $4/3$ denotes the Casimir factor and $g$ represents the 
running coupling in the presence of a strong magnetic field 
\cite{Ferrer:PRD91'2015},
\begin{eqnarray}
g^2 &=& \frac{4\pi}{{\alpha_s^0(\mu_0)}^{-1}+\frac{11N_c}{12\pi}
\ln\left(\frac{\Lambda_{QCD}^2+M^2_B}{\mu_0^2}\right)+\frac{1}{3\pi}\sum_i\frac{|q_i B|}{\tau}}
~,\end{eqnarray}
where
\begin{eqnarray}
\alpha_s^0(\mu_0) &=& \frac{12\pi}
{11N_c\ln\left(\frac{\mu_0^2+M^2_B}{\Lambda_V^2}\right)}
,\end{eqnarray}
where $M_B$ ($\sim 1$ GeV) represents an infrared mass 
which is interpreted as the ground state mass of two 
gluons connected by a fundamental string, with the string tension, 
$\tau=0.18 ~{\rm{GeV}}^2$, and $\Lambda_V$ and $\mu_0$ have values 
$0.385$ GeV and $1.1$ GeV, respectively \cite{Ferrer:PRD91'2015, Simonov:PAN58'1995,Andreichikov:PRL110'2013}.

$S(k)$ is the quark propagator, which in the strong magnetic 
field limit is given \cite{Schwinger:PR82'1951} by the 
Schwinger proper-time method in the momentum space,
\be\label{q. propagator}
S(k)=ie^{-\frac{k^2_\perp}{|q_iB|}}\frac{\left(\gamma^0 k_0-\gamma^3 k_z+m_i\right)}{k^2_\parallel-m^2_i}\left(1
-\gamma^0\gamma^3\gamma^5\right)
,\ee
where the four vectors are defined with the metric tensors: 
$g^{\mu\nu}_\perp={\rm{diag}}(0,-1,-1,0)$ and $g^{\mu\nu}_\parallel= 
{\rm{diag}}(1,0,0,-1)$,
\begin{eqnarray*}
&& k_{\perp\mu}\equiv(0,k_x,k_y,0), ~~ k_{\parallel\mu}\equiv(k_0,0,0,k_z)
~.\end{eqnarray*}
$D^{\mu \nu}(p-k)$ is the gluon propagator, which is not affected 
by the magnetic field, {\em i.e.},
\be
\label{g. propagator}
D^{\mu \nu} (p-k)=\frac{ig^{\mu \nu}}{(p-k)^2}
~.\ee
In imaginary-time formalism, the quark self-energy \eqref{Q.S.E.} 
in strong magnetic field can be simplified \cite{Conductivities} into
\begin{eqnarray}\label{Q.S.E.(3)}
\Sigma(p_\parallel)=\frac{g^2|q_iB|}{3\pi^2}\left[\frac{\pi T}{2m_i}-\ln(2)\right]\left[\frac{\gamma^0p_0}{p_\parallel^2}+\frac{\gamma^3p_z}{p_\parallel^2}+\frac{\gamma^0\gamma^5p_z}{p_\parallel^2}+\frac{\gamma^3\gamma^5p_0}{p_\parallel^2}\right]
.\end{eqnarray}

To solve the Schwinger-Dyson equation self-consistently, the 
quark self-energy at finite temperature in the presence of 
magnetic field should be written first in a covariant form~\cite{Ayala:PRD91'2015,Karmakar:PRD99'2019},
\begin{equation}\label{1general q.s.e.}
\Sigma(p_\parallel)=A\gamma^\mu u_\mu+B\gamma^\mu b_\mu+C\gamma^5\gamma^\mu u_\mu+D\gamma^5\gamma^\mu b_\mu
~,\end{equation}
where the form factors, $A$, $B$, $C$ and $D$ are computed in LLL 
approximation as
\begin{eqnarray}
&&A=\frac{g^2|q_iB|}{3\pi^2}\left[\frac{\pi T}{2m_i}-\ln(2)\right]\frac{p_0}{p_\parallel^2} ~, \\ 
&&B=\frac{g^2|q_iB|}{3\pi^2}\left[\frac{\pi T}{2m_i}-\ln(2)\right]\frac{p_z}{p_\parallel^2} ~, \\ 
&&C=-\frac{g^2|q_iB|}{3\pi^2}\left[\frac{\pi T}{2m_i}-\ln(2)\right]\frac{p_z}{p_\parallel^2} ~, \\ 
&&D=-\frac{g^2|q_iB|}{3\pi^2}\left[\frac{\pi T}{2m_i}-\ln(2)\right]\frac{p_0}{p_\parallel^2}
~,\end{eqnarray}
with $u^\mu$ (1,0,0,0) and $b^\mu$ (0,0,0,-1), the preferred directions 
of heat bath and magnetic field, respectively.

The quark self-energy \eqref{1general q.s.e.} can be expressed in terms 
of chiral projection operators ($P_R$ and $P_L$) as
\begin{equation}\label{projection1}
\Sigma(p_\parallel)=P_R\left[(A-B)\gamma^\mu u_\mu+(B-A)\gamma^\mu b_\mu
\right]P_L+P_L\left[(A+B)\gamma^\mu u_\mu+(B+A)\gamma^\mu b_\mu\right]P_R
~,\end{equation}
after substituting $C=-B$ and $D=-A$. Hence, the Schwinger-Dyson equation 
\eqref{S.D.E.} finally (in appendix \ref{A.T.M.}) gives the thermal mass 
for $i$-th flavor (through the $p_0=0, p_z \rightarrow 0$ limit) in 
a strong magnetic field as
\begin{eqnarray}\label{Mass}
m_{iT,B}^2 (T,B) =\frac{g^2|q_iB|}{3\pi^2}\left[\frac{\pi T}{2m_i}-\ln(2)\right]
,\end{eqnarray}
which depends on both temperature and magnetic field. Thus the gluon and 
quark distribution functions with medium generated 
masses (\ref{Gluon mass},\ref{Gluon mass(eb)}) and 
(\ref{Quark mass},\ref{Mass}) for gluons and quarks, respectively 
manifest the interactions present in the medium in terms of modified 
occupation probabilities in the phase space, which in turn affect the 
transport coefficients related to the momentum transport 
in kinetic theory approach in next section.

\section{Momentum transport in a thermal QCD medium}
In this section, we will study the transport coefficients for a strongly
interacting matter through the shear and bulk viscosities in the presence of 
momentum anisotropies. The shear and bulk viscosities can be determined using 
different models and approaches, {\em namely} the relativistic Boltzmann transport 
equation in the relaxation-time approximation \cite{Danielewicz:PRD31'1985,Heckmann:EPJA48'2012,Yasui:PRD96'2017}, the 
correlator technique using Green-Kubo formula \cite{Basagoiti:PRD66'2002,Kharzeev:JHEP0809'2008,
Moore:JHEP0809'2008,Plumari:PRC86'2012}, the lattice simulations \cite{Astrakhantsev:JHEP1704'2017,Astrakhantsev:PRD98'2018}, the 
molecular dynamics simulation \cite{Gelman:PRC74'2006} etc. In the 
present analysis, we use the relativistic Boltzmann transport 
equation to calculate the shear and bulk viscosities 
in the relaxation-time approximation for both isotropic and anisotropic hot 
QCD mediums in subsections 3.1 and 3.2, respectively.

\subsection{Shear and bulk viscosities for an isotropic thermal medium}
To proceed for the calculations of shear and bulk viscosities, we assume a 
local temperature $T(x)$ and flow velocity $u^\mu(x)$ which is 
also called as the velocity of energy transport in the 
Landau-Lifshitz approach and the velocity of baryon number 
flow in the Eckart approach. In this work, we assume the baryon 
chemical potential to be very small or zero.

Allowing the system to be slightly out of equilibrium, the 
energy-momentum tensor gets shifted by a small amount, {\em i.e.},
\begin{eqnarray}
\Delta T^{\mu\nu}=T^{\mu\nu}-T_{(0)}^{\mu\nu}
~,\end{eqnarray}
where $T_{(0)}^{\mu\nu}$ represents the energy-momentum tensor 
in local equilibrium and $T^{\mu\nu}$ for the partonic system 
is given by
\be
T^{\mu\nu}=\int\frac{d^3{\rm p}}{(2\pi)^3}p^\mu p^\nu \left[2\sum_i g_i\frac{f_i}{\omega_i}+g_g\frac{f_g}{\omega_g}\right]
,\ee
where the factor ``2'' represents the equal contributions 
from quark and antiquark. The nonequilibrium part of the 
energy-momentum tensor is proportional to the velocity 
gradient. The traceless part and the trace part of the 
velocity gradient are known as the shear viscous force 
and the bulk viscous force, respectively.
\be\label{em1}
\Delta T^{\mu\nu}=\int\frac{d^3{\rm p}}{(2\pi)^3}p^\mu p^\nu \left[2\sum_i g_i\frac{\delta f_i}{\omega_i}+g_g\frac{\delta f_g}{\omega_g}\right]
,\ee
where the summation is over three light flavors ($u$, $d$ and $s$) and 
$g_i$ and $g_g$ are the degeneracy factors for quarks and gluons, 
respectively. The infinitesimal change in quark distribution function 
due to the action of an external force is defined as 
$\delta f_i=f_i-f_i^{\rm iso}$, where $f_i^{\rm iso}$ 
is the equilibrium distribution function in the isotropic medium for $i$th flavor, 
\be\label{Q.D.F.}
f_i^{\rm iso}=\frac{1}{e^{\beta u^\alpha p_\alpha}+1}
~,\ee
where $p_\alpha\equiv\left(\omega_i,\mathbf{p}\right)$ with  $\omega_i=
\sqrt{\mathbf{p}^2+m_i^2}$ and $u^\alpha$ is the four-velocity of fluid. 
Similarly, the infinitesimal change in gluon distribution 
function is defined as $\delta f_g=f_g-f_g^{\rm iso}$, 
where $f_g^{\rm iso}$ is the equilibrium distribution 
function in the isotropic medium,
\be\label{G.D.F.}
f_g^{\rm iso}=\frac{1}{e^{\beta u^\alpha p_\alpha}-1}
~,\ee
with $p_\alpha\equiv\left(\omega_g,\mathbf{p}\right)$.
The infinitesimal changes in the distribution functions for 
gluons and quarks can be obtained from the solutions of their
respective relativistic Boltzmann transport equations. It
will be easier to solve in the relaxation-time approximation: 
\begin{eqnarray}
&&p^\mu\partial_\mu f_g(x,p) = -\frac{p_\nu u^\nu}{\tau_g}\delta f_g(x,p),\\
&&p^\mu\partial_\mu f_i(x,p) = -\frac{p_\nu u^\nu}{\tau_i}\delta f_i(x,p),
\end{eqnarray}
where the forms of the relaxation times for gluons ($\tau_g$) and quarks 
($\tau_i$) can be understood heuristically in terms 
of the quasiparticle description in section 2, where they 
acquire masses due to the interactions among themselves in a thermal QCD medium.
Let us start with the relaxation time in the case of pure SU(3) gauge 
theory and then extend to the case where the quarks are included: \\ 
The gluon-gluon interaction exhibits the infrared singularities when the 
momentum of an exchanged gluon becomes soft, at least, in the naive 
perturbation theory, because the gluons are massless.
This is circumvented by using a resummed (dressed) gluon propagator 
in a thermal medium, which 
is decomposed into the longitudinal, $\Delta_L$ and transverse, $\Delta_T$ 
components. The longitudinal one in the static limit manifests the gluon to 
acquire an effective mass, {\em namely} 
\begin{eqnarray}
\Delta_L (0, {\bf q})=\frac{1}{{\bf q}^2+2 m_{gT}^2},
\end{eqnarray}
where the effective (thermal) mass (given in eq. \eqref{Gluon mass}), 
in turn, screens the infrared singularities, known as familiar Debye 
screening. Whereas the transverse one, 
$\Delta_T (0, \bf q)$ (=$\frac{1}{\bf q^2}$), at first sight, 
implies that the magnetostatic fields 
are not screened. However, if the leading term in $q_0/{\bf q}$ is
retained, then it yields for $ q_0/{\bf q} \rightarrow 0$,
\begin{eqnarray}
\Delta_T (0, {\bf q}) \simeq \frac{1}{{\bf q}^2 + \frac{i}{2} \pi m^2_{gT} (q_0/{\bf q})},
\label{deltaT}
\end{eqnarray}
showing a frequency-dependent (dynamical) screening with a cut-off, 
which is able to screen the infrared singularities to make
the cross-sections finite, otherwise those cross-sections would 
diverge in the bare perturbation theory. So, the gluon-gluon 
cross-section is being computed with dressed gluon propagator, thus
the cross-section consists of $|\Delta_L|^2$, $|\Delta_T|^2$ and their 
interference term, where the first one is made finite  
by the Debye screening. Both the second and the interference terms are 
made finite in HTL approximation and one recovers, as in the 
case of Debye screening, a $\ln (T/m_{\rm gT})$ screening factor. This 
factor is not affected by the possible existence of a magnetic 
mass, demonstrating that despite the absence of the 
screening of magnetostatic fields, transverse gluon exchange is 
effectively cut off in the infrared by the thermal mass. Thus, the 
(quasiparticle) interactions also play the role in deriving the relaxation 
time for gluons, which is of the order of $\tau_g \sim \left[\alpha_s^2 
\ln (1/\alpha_s)\right]^{-1}$ \cite{Hosoya:NPB250'1985,Baym:PRL64'1990}. 

The preceding discussion can easily be generalized to include the quarks,
where the infrared singularities in the relevant processes ($gg \rightarrow gg, 
qg \rightarrow qg,qq \rightarrow qq$) responsible to bring back the system 
into local equilibrium are similarly removed by the masses generated by a 
thermal medium. The thermal masses for light quarks, $m^2_{qT}$ (given in 
eq. \eqref{Quark mass}) in hard thermal loop calculation are independent 
of their masses and are of the same order of $m^2_{gT}$, apart from a 
flavor factor. One then finds that the 
gluon and quark contributions are simply added to yield the 
final form of the relaxation time. Indeed, the explicit expressions 
for the relaxation times of gluons ($\tau_g$) and quarks ($\tau_i$) are 
calculated in ref. \cite{Hosoya:NPB250'1985}, 
\begin{eqnarray}
&&\tau_g=\frac{1}{22.5T\alpha_s^2\log\left(1/\alpha_s\right)\left[1+0.06N_i
\right]} ~, \label{G.R.T.} \\
&&\tau_i=\frac{1}{5.1T\alpha_s^2\log\left(1/\alpha_s\right)\left[1+0.12 
(2N_i+1)\right]} \label{Q.R.T.} 
~,\end{eqnarray}
respectively. However, in the heavy quark transport phenomena, if the 
heavy quarks are assumed to be equilibrated in the medium, one can 
define the relaxation time for heavy quark, which carries  the mass 
dependence. 

Substituting the values of $\delta f_i$ and 
$\delta f_g$ in eq. \eqref{em1}, we obtain
\be\label{em2}
\Delta T^{\mu\nu}=-\int\frac{d^3{\rm p}}{(2\pi)^3}\frac{p^\mu p^\nu}{p_\nu u^\nu} \left[2\sum_i g_i\frac{\tau_i p^\mu\partial_\mu f_i}{\omega_i}+g_g\frac{\tau_g p^\mu\partial_\mu f_g}{\omega_g}\right]
.\ee
The derivative is written covariantly as the sum of the time and 
space parts: $\partial_\mu=u_\mu D+\nabla_\mu$, with 
$D=u^\mu\partial_\mu$. In the local rest frame, the flow 
velocity and the temperature are the functions of spatial and temporal 
coordinates, so the distribution function can be expanded in terms 
of the gradients of flow velocity and temperature. The partial 
derivatives of the isotropic quark and gluon distribution 
functions are calculated as
\be
&&\partial_\mu f_i^{\rm iso}=\frac{f_i^{\rm iso}(1-f_i^{\rm iso})}{T}\left[u_\alpha p^\alpha u_\mu\frac{DT}{T}+u_\alpha p^\alpha\frac{\nabla_\mu T}{T}-u_\mu p^\alpha Du_\alpha-p^\alpha\nabla_\mu u_\alpha\right] , \\ 
&&\partial_\mu f_g^{\rm iso}=\frac{f_g^{\rm iso}(1+f_g^{\rm iso})}{T}\left[u_\alpha p^\alpha u_\mu\frac{DT}{T}+u_\alpha p^\alpha\frac{\nabla_\mu T}{T}-u_\mu p^\alpha Du_\alpha-p^\alpha\nabla_\mu u_\alpha\right]
,\ee
respectively. Substituting the above values of $\partial_\mu f_i^{\rm iso}$ and 
$\partial_\mu f_g^{\rm iso}$ in eq. \eqref{em2}, then using 
$\frac{DT}{T}=-\left(\frac{\partial P}{\partial \varepsilon}\right)\nabla_\alpha u^\alpha$ and $Du_\alpha=\frac{\nabla_\alpha P}{\varepsilon+P}$ from the 
energy-momentum conservation, we get
\be\label{em3}
\nonumber\Delta T^{\mu\nu} &=& 2\sum_i g_i\int\frac{d^3{\rm p}}{(2\pi)^3}\frac{p^\mu p^\nu}{\omega_i T} ~ \tau_i ~ f_i^{\rm iso}(1-f_i^{\rm iso})\left[\omega_i\left(\frac{\partial P}{\partial \varepsilon}\right)\nabla_\alpha u^\alpha+p^\alpha\left\lbrace\frac{\nabla_\alpha P}{\varepsilon+P}-\frac{\nabla_\alpha T}{T}\right\rbrace\right. \\ && \left.\nonumber+\frac{p^\alpha p^\beta}{\omega_i}\nabla_\alpha u_\beta\right]+g_g\int\frac{d^3{\rm p}}{(2\pi)^3}\frac{p^\mu p^\nu}{\omega_g T} ~ \tau_g ~ f_g^{\rm iso}(1+f_g^{\rm iso})\left[\omega_g\left(\frac{\partial P}{\partial \varepsilon}\right)\nabla_\alpha u^\alpha\right. \\ && \left.+p^\alpha\left\lbrace\frac{\nabla_\alpha P}{\varepsilon+P}-\frac{\nabla_\alpha T}{T}\right\rbrace+\frac{p^\alpha p^\beta}{\omega_g}\nabla_\alpha u_\beta\right]
.\ee
The pressure and the energy density are related to the 
energy-momentum tensor as 
$P=-\Delta_{\mu\nu}T^{\mu\nu}/3$ and 
$\varepsilon=u_\mu T^{\mu\nu}u_\nu$, where the projection tensor 
is defined as $\Delta_{\mu\nu}=g_{\mu\nu}-u_\mu u_\nu$. The 
definitions of viscosities require the velocity gradient to 
be nonzero. The freedom to define velocity $u^\mu$ or, 
equivalently, the local rest frame creates arbitrariness, 
because in the Eckart frame $u^\mu$ represents the velocity of 
baryon number flow, whereas in the Landau-Lifshitz frame it 
represents the velocity of energy flow. However, the arbitrariness 
can be avoided by choosing a specific frame through the imposition 
of the ``condition of fit". To choose the Landau-Lifshitz frame, the 
condition of fit in the local rest frame requires the $``00"$ 
component of the dissipative part of the energy-momentum tensor to 
be zero, {\em i.e.}, $\Delta T^{00}=0$ \cite{Albright:PRC93'2016}. 
Since our motivation is to calculate shear and bulk 
viscosities, we write only the space-space component of 
$\Delta T^{\mu\nu}$ which is proportional to the velocity gradient,
\be\label{em4}
\nonumber\Delta T^{ij} &=& 2\sum_i g_i\int\frac{d^3{\rm p}}{(2\pi)^3}\frac{p^i p^j}{\omega_i T} ~ \tau_i ~ f_i^{\rm iso}(1-f_i^{\rm iso})\left[-\frac{p^kp^l}{2\omega_i}W_{kl}+\left\lbrace \omega_i\left(\frac{\partial P}{\partial \varepsilon}\right)-\frac{\rm p^2}{3\omega_i}\right\rbrace\partial_l u^l\right. \\ && \left.\nonumber + p^k\left\lbrace \frac{\partial_k P}{\varepsilon+P}-\frac{\partial_k T}{T} \right\rbrace\right]+g_g\int\frac{d^3{\rm p}}{(2\pi)^3}\frac{p^i p^j}{\omega_g T} ~ \tau_g ~ f_g^{\rm iso}(1+f_g^{\rm iso})\left[-\frac{p^kp^l}{2\omega_g}W_{kl}\right. \\ && \left.+\left\lbrace \omega_g\left(\frac{\partial P}{\partial \varepsilon}\right)-\frac{\rm p^2}{3\omega_g}\right\rbrace\partial_l u^l+p^k\left\lbrace \frac{\partial_k P}{\varepsilon+P}-\frac{\partial_k T}{T} \right\rbrace\right]
,\ee
where the following expressions have been used :
\be
\partial_k u_l &=&-\frac{1}{2}W_{kl}-\frac{1}{3}\delta_{kl}\partial_j u^j, \\
W_{kl} &=& \partial_k u_l+\partial_l u_k-\frac{2}{3}\delta_{kl}\partial_j u^j
.\ee

In a fluid, fluctuations in the momentum and energy densities 
represent two of the hydrodynamic modes whose responses are 
characterized by the shear viscosity ($\eta$) and the bulk 
viscosity ($\zeta$), respectively. For the system which is 
slightly shifted from the equilibrium, the shear and bulk 
viscosities are defined as the coefficients of the 
space-space component of the dissipative part of the 
energy-momentum tensor in a first order theory \cite{Lifshitz:BOOK'1981,Hosoya:NPB250'1985,Landau:BOOK'1987}, 
\be\label{definition}
\Delta T^{ij}=-\eta W^{ij}-\zeta\delta^{ij}\partial_l u^l
.\ee
This relation is valid for small fluctuations of the 
energy-momentum tensor from its equilibrium. We get 
the shear viscosity and the bulk viscosity by 
comparing equations \eqref{em4} and 
\eqref{definition} for an isotropic medium as
\begin{eqnarray}\label{iso.eta}
\eta^{\rm iso}=\frac{\beta}{15\pi^2}\sum_i g_i \int d{\rm p}~\frac{{\rm p}^6}{\omega_i^2} ~ \tau_i ~ f_i^{\rm iso}(1-f_i^{\rm iso})+\frac{\beta}{30\pi^2} g_g \int d{\rm p}~\frac{{\rm p}^6}{\omega_g^2} ~ \tau_g ~ f_g^{\rm iso}(1+f_g^{\rm iso})
~,\end{eqnarray}
\begin{eqnarray}\label{iso.zeta1}
\zeta^{\rm iso}=\frac{2}{3}\sum_i g_i \int\frac{d^3{\rm p}}{(2\pi)^3}~\frac{{\rm p}^2}{\omega_i} ~ f_i^{\rm iso}(1-f_i^{\rm iso})A_i+\frac{1}{3}g_g \int\frac{d^3{\rm p}}{(2\pi)^3}~\frac{{\rm p}^2}{\omega_g} ~ f_g^{\rm iso}(1+f_g^{\rm iso})A_g
~.\end{eqnarray}
The factors $A_i$ and $A_g$ in the $\zeta^{\rm iso}$ expression are given by
\begin{eqnarray}
&&A_i=\frac{\tau_i}{3T}\left[\frac{{\rm p}^2}{\omega_i}-3\left(\frac{\partial P}{\partial \varepsilon}\right)\omega_i\right],\label{Ai} \\ 
&&A_g=\frac{\tau_g}{3T}\left[\frac{{\rm p}^2}{\omega_g}-3\left(\frac{\partial P}{\partial \varepsilon}\right)\omega_g\right]\label{Ag}
.\end{eqnarray}
For the calculation of bulk viscosity, the forms of $A_i$ and $A_g$ 
should be such that, the Landau-Lifshitz condition, {\em i.e.} 
$u_\mu \Delta T^{\mu\nu}u_\nu=0$ is satisfied. In the local rest 
frame, to make the Landau-Lifshitz condition ($\Delta T^{00}=0$) 
satisfied, we have to replace $A_i\rightarrow A_i^\prime=A_i-b_i\omega_i$ and 
$A_g\rightarrow A_g^\prime=A_g-b_g\omega_g$, where $b_i$ and $b_g$ 
are associated with the energy conservation 
\cite{Chakraborty:PRC83'2011}. From eq. \eqref{em3}, the 
Landau-Lifshitz conditions for terms $A_i$ and $A_g$ are written as
\begin{eqnarray}
&&2\sum_i g_i\int\frac{d^3{\rm p}}{(2\pi)^3}~\omega_i f_i^{\rm iso}(1-f_i^{\rm iso})\left(A_i-b_i\omega_i\right)=0 \label{A_i} ~,~ \\ 
&&g_g\int\frac{d^3{\rm p}}{(2\pi)^3}~\omega_g f_g^{\rm iso}(1+f_g^{\rm iso})\left(A_g-b_g\omega_g\right)=0 \label{A_g}
~,\end{eqnarray}
respectively, and the quantities $b_i$ and $b_g$ are obtained 
by solving equations \eqref{A_i} and \eqref{A_g}. Now 
replacing 
$A_i\rightarrow A_i^\prime$ and $A_g\rightarrow A_g^\prime$ 
in eq. \eqref{iso.zeta1} and then simplifying, we get the 
bulk viscosity for an isotropic medium as
\begin{eqnarray}\label{iso.zeta}
\nonumber\zeta^{\rm iso} &=& \frac{\beta}{9\pi^2}\sum_i g_i \int d{\rm p}~{\rm p}^2\left[\frac{{\rm p}^2}{\omega_i}-3\left(\frac{\partial P}{\partial \varepsilon}\right)\omega_i\right]^2 \tau_i ~ f_i^{\rm iso}(1-f_i^{\rm iso}) \\ && +\frac{\beta}{18\pi^2}g_g\int d{\rm p}~{\rm p}^2\left[\frac{{\rm p}^2}{\omega_g}-3\left(\frac{\partial P}{\partial \varepsilon}\right)\omega_g\right]^2 \tau_g ~ f_g^{\rm iso}(1+f_g^{\rm iso})
~.\end{eqnarray}

\subsection{Shear and bulk viscosities for an anisotropic thermal medium}
Here we are going to study the shear and bulk viscosities 
in two different types of momentum anisotropies, which 
may be produced at very early stages of ultrarelativistic 
heavy ion collisions. The first one is due to the initial 
asymptotic expansion and the second one is due to the strong 
magnetic field.

\subsubsection{Expansion-induced anisotropy}
The QGP created in the early stages of heavy ion collisions 
experiences larger longitudinal expansion than the radial 
expansion which develops a local momentum anisotropy. If the 
momentum anisotropy is weak ($\xi<1$) with direction $\mathbf{n}$, the 
distribution function in anisotropic 
medium can be approximated as the isotropic one with the tail 
of distribution being curtailed \cite{Romatschke:PRD68'2003}. The 
distribution function is thus rescaled as $f_{{\rm ex},i}^{\rm aniso}(\mathbf{p})=f_{i}^{\rm iso}(\sqrt{\mathbf{p}^2+\xi(\mathbf{p}\cdot\mathbf{n})^2})$, {\em i.e.},
\be\label{A.D.F.}
f_{{\rm ex},i}^{\rm aniso}(\mathbf{p};T)=\frac{1}{e^{\beta\sqrt{\rm{p}^2+\xi(\mathbf{p}\cdot\mathbf{n})^2+m_i^2}}+1}
~,\ee
which after Taylor series expansion up to 
$\mathcal{O}(\xi)$, takes the following form,
\be\label{expansion}
f_{{\rm ex},i}^{\rm aniso}=f_i^{\rm iso}-\frac{\xi\beta(\mathbf{p}\cdot\mathbf{n})^2}{2\omega_i}f_i^{\rm iso}(1-f_i^{\rm iso})
~.\ee
Similarly the anisotropic distribution function for gluon is written as
\be\label{expansion_gluon}
f_{{\rm ex},g}^{\rm aniso}=f_g^{\rm iso}-\frac{\xi\beta(\mathbf{p}\cdot\mathbf{n})^2}{2\omega_g}f_g^{\rm iso}(1+f_g^{\rm iso})
~.\ee
The general form of the anisotropic parameter ($\xi$) is written as
\be\label{parameter}
\xi=\frac{\left\langle\mathbf{p}_T^2\right\rangle}{2\left\langle p_L^2\right\rangle}-1
~,\ee
where $p_L=\mathbf{p}\cdot\mathbf{n}$, $\mathbf{p}_T=\mathbf{p}-\mathbf{n}\cdot(\mathbf{p}\cdot\mathbf{n})$, $\mathbf{p}\equiv(\rm{p}\sin\theta\cos\phi,\rm{p}\sin\theta\sin\phi,\rm{p}\cos\theta)$, $\mathbf{n}=(\sin\alpha,0,\cos\alpha)$, 
$\alpha$ is the angle between z-axis and direction of anisotropy, and  $(\mathbf{p}\cdot\mathbf{n})^2=\rm{p}^2c(\alpha,\theta,\phi)=\rm{p}^2(\sin^2\alpha\sin^2\theta\cos^2\phi+\cos^2\alpha\cos^2\theta
+\sin(2\alpha)\sin\theta\cos\theta\cos\phi)$. For $p_T\gg p_L$, $\xi$ is positive. 

In the presence of weak-momentum anisotropy, the partial 
derivatives of the anisotropic quark and gluon distribution 
functions are calculated as
\be
\nonumber\partial_\mu f_{{\rm ex},i}^{\rm aniso} &=& \partial_\mu f_i^{\rm iso}-\frac{\xi{\rm p}^2c(\theta , \phi)}{2}\left[-\frac{f_i^{\rm iso}(1-f_i^{\rm iso})}{\omega_i T^2}\left(u_\mu DT+\nabla_\mu T\right)\right. \\ && \left.-\frac{f_i^{\rm iso}(1-f_i^{\rm iso})}{\omega_i^2 T}\left(u_\mu p_\alpha Du^\alpha+p_\alpha\nabla_\mu u^\alpha\right)+\frac{1-2f_i^{\rm iso}}{\omega_i T}\partial_\mu f_i^{\rm iso}\right]
,\ee
\be
\nonumber\partial_\mu f_{{\rm ex},g}^{\rm aniso} &=& \partial_\mu f_g^{\rm iso}-\frac{\xi{\rm p}^2c(\theta , \phi)}{2}\left[-\frac{f_g^{\rm iso}(1+f_g^{\rm iso})}{\omega_g T^2}\left(u_\mu DT+\nabla_\mu T\right)\right. \\ && \left.-\frac{f_g^{\rm iso}(1+f_g^{\rm iso})}{\omega_g^2 T}\left(u_\mu p_\alpha Du^\alpha+p_\alpha\nabla_\mu u^\alpha\right)+\frac{1+2f_g^{\rm iso}}{\omega_g T}\partial_\mu f_g^{\rm iso}\right]
,\ee
respectively. Now substituting $\partial_\mu f_{{\rm ex},i}^{\rm aniso}$ 
and $\partial_\mu f_{{\rm ex},g}^{\rm aniso}$ in eq. \eqref{em2} for the 
expansion-driven anisotropy and then proceeding like the isotropic case, 
we obtain the shear and bulk viscosities as follows,
\be\label{A.S.V.}
\nonumber\eta_{\rm ex}^{\rm aniso} &=& \frac{\beta}{15\pi^2}\sum_i g_i \int d{\rm p}~\frac{{\rm p}^6}{\omega_i^2} ~ \tau_i ~ f_i^{\rm iso}(1-f_i^{\rm iso})-\frac{\xi\beta}{90\pi^2}\sum_i g_i \int d{\rm p}\frac{{\rm p}^8}{\omega_i^4} ~ \tau_i ~ f_i^{\rm iso}(1-f_i^{\rm iso}) \\ && \nonumber-\frac{\xi\beta^2}{90\pi^2}\sum_i g_i \int d{\rm p}\frac{{\rm p}^8}{\omega_i^3} ~ \tau_i ~ f_i^{\rm iso}(1-f_i^{\rm iso})(1-2f_i^{\rm iso}) \\ && \nonumber+\frac{\beta}{30\pi^2} g_g \int d{\rm p}~\frac{{\rm p}^6}{\omega_g^2} ~ \tau_g ~ f_g^{\rm iso}(1+f_g^{\rm iso})-\frac{\xi\beta}{180\pi^2} g_g \int d{\rm p}\frac{{\rm p}^8}{\omega_g^4} ~ \tau_g ~ f_g^{\rm iso}(1+f_g^{\rm iso}) \\ && -\frac{\xi\beta^2}{180\pi^2} g_g \int d{\rm p}\frac{{\rm p}^8}{\omega_g^3} ~ \tau_g ~ f_g^{\rm iso}(1+f_g^{\rm iso})(1+2f_g^{\rm iso})
~,\ee
where the $\xi$-independent terms in right hand side 
constitute the shear viscosity for an isotropic 
medium. So in terms of $\eta^{\rm iso}$, 
$\eta_{\rm ex}^{\rm aniso}$ is written as
\be\label{A.S.V.(1)}
\nonumber\eta_{\rm ex}^{\rm aniso} &=& \eta^{\rm iso} - \xi\left[\frac{\beta^2}{90\pi^2}\sum_i g_i \int d{\rm p}\frac{{\rm p}^8}{\omega_i^3}~ \tau_i ~ f_i^{\rm iso}(1-f_i^{\rm iso})\left\lbrace \frac{1}{\beta \omega_i}+1-2f_i^{\rm iso} \right\rbrace\right. \\ && \left.+\frac{\beta^2}{180\pi^2} g_g \int d{\rm p}\frac{{\rm p}^8}{\omega_g^3}~ \tau_g ~ f_g^{\rm iso}(1+f_g^{\rm iso})\left\lbrace \frac{1}{\beta \omega_g}+1+2f_g^{\rm iso} \right\rbrace\right]
.\ee
The bulk viscosity is calculated as
\be\label{A.B.V.}
\nonumber\zeta_{\rm ex}^{\rm aniso} &=& \frac{\beta}{9\pi^2}\sum_i g_i \int d{\rm p}~{\rm p}^2 \left[\frac{{\rm p}^2}{\omega_i}-3\left(\frac{\partial P}{\partial \varepsilon}\right)\omega_i\right]^2 \tau_i ~ f_i^{\rm iso}(1-f_i^{\rm iso}) \\ && \nonumber -\frac{\xi\beta}{54\pi^2}\sum_i g_i \int d{\rm p}\frac{{\rm p}^4}{\omega_i^2}\left[\frac{{\rm p}^4}{\omega_i^2}-9\left(\frac{\partial P}{\partial \varepsilon}\right)^2\omega_i^2\right] \tau_i ~ f_i^{\rm iso}(1-f_i^{\rm iso}) \\ && \nonumber-\frac{\xi\beta^2}{54\pi^2}\sum_i g_i \int d{\rm p}\frac{{\rm p}^4}{\omega_i}\left[\frac{{\rm p}^2}{\omega_i}-3\left(\frac{\partial P}{\partial \varepsilon}\right)\omega_i\right]^2 \tau_i ~ f_i^{\rm iso}(1-f_i^{\rm iso})(1-2f_i^{\rm iso}) \\ && \nonumber+\frac{\beta}{18\pi^2} g_g \int d{\rm p}~{\rm p}^2\left[\frac{{\rm p}^2}{\omega_g}-3\left(\frac{\partial P}{\partial \varepsilon}\right)\omega_g\right]^2 \tau_g ~ f_g^{\rm iso}(1+f_g^{\rm iso}) \\ && \nonumber -\frac{\xi\beta}{108\pi^2} g_g \int d{\rm p}\frac{{\rm p}^4}{\omega_g^2}\left[\frac{{\rm p}^4}{\omega_g^2}-9\left(\frac{\partial P}{\partial \varepsilon}\right)^2\omega_g^2\right] \tau_g ~ f_g^{\rm iso}(1+f_g^{\rm iso}) \\ && -\frac{\xi\beta^2}{108\pi^2} g_g \int d{\rm p}\frac{{\rm p}^4}{\omega_g}\left[\frac{{\rm p}^2}{\omega_g}-3\left(\frac{\partial P}{\partial \varepsilon}\right)\omega_g\right]^2 \tau_g ~ f_g^{\rm iso}(1+f_g^{\rm iso})(1+2f_g^{\rm iso})
~,\ee
which can be decomposed into $\xi$-independent (isotropic) and 
$\xi$-dependent parts as
\be\label{A.B.V.(1)}
\nonumber\zeta_{\rm ex}^{\rm aniso} &=& \zeta_{\rm ex}^{\rm iso}-\xi\left[\frac{\beta^2}{54\pi^2}\sum_i g_i \int d{\rm p}\frac{{\rm p}^4}{\omega_i} ~ \tau_i ~ f_i^{\rm iso}(1-f_i^{\rm iso})\left\lbrace\frac{1}{\beta\omega_i}\left[\frac{{\rm p}^4}{\omega_i^2}-9\left(\frac{\partial P}{\partial \varepsilon}\right)^2\omega_i^2\right] \right.\right. \\ && \left.\left.\nonumber+(1-2f_i^{\rm iso})\left[\frac{{\rm p}^2}{\omega_i}-3\left(\frac{\partial P}{\partial \varepsilon}\right)\omega_i\right]^2\right\rbrace+\frac{\beta^2}{108\pi^2} g_g \int d{\rm p}\frac{{\rm p}^4}{\omega_g} ~ \tau_g ~ f_g^{\rm iso}(1+f_g^{\rm iso})\right. \\ && \left.\times\left\lbrace\frac{1}{\beta\omega_g}\left[\frac{{\rm p}^4}{\omega_g^2}-9\left(\frac{\partial P}{\partial \varepsilon}\right)^2\omega_g^2\right]+(1+2f_g^{\rm iso})\left[\frac{{\rm p}^2}{\omega_g}-3\left(\frac{\partial P}{\partial \varepsilon}\right)\omega_g\right]^2\right\rbrace\right]
.\ee

\subsubsection{Strong magnetic field-induced anisotropy}
The presence of magnetic field makes the quark momentum 
$\mathbf{p}$ to decompose into the transverse and 
longitudinal components with respect to its direction 
(say, $3$-direction). Thus the dispersion relation for the 
quark of $i$th flavor is modified as
\begin{eqnarray}\label{dispersion relation}
\omega_{i,n}(p_L)=\sqrt{p_L^2+2n\left|q_iB\right|+m_i^2}
~,\end{eqnarray}
where $n=0$, $1$, $2$, $\cdots$ specify different Landau levels. 
In the strong magnetic field limit, the strength of the magnetic 
field is much larger than the temperature of the system and the 
mass of the quark. So, even in a thermal medium the quarks can not 
get excited to higher Landau levels due to very high energy gap 
$\sim \mathcal{O}(\sqrt{eB})$ and they occupy only the 
lowest Landau level. Therefore, $p_T$ is much smaller than $p_L$ 
and this develops a momentum anisotropy with the value of the 
anisotropic parameter ($\xi$) becomes negative. The distribution 
function in this case has the following form,
\be\label{A.D.F.(eB)}
f_{{\rm B},i}^{\rm aniso}(\mathbf{p^\prime};T)=\frac{1}{e^{\beta\sqrt{{{\rm p}^\prime}^2+\xi(\mathbf{p^\prime}\cdot\mathbf{n})^2+m_i^2}}+1}
~,\ee
where we have denoted the momentum vector in strong magnetic field limit 
($p_T =0$) by $\mathbf{p^\prime}=(0,0,p_3)$. For very small $\xi$, the above 
distribution function can be expanded as
\be
f_{{\rm B},i}^{\rm aniso}=f_i^{\xi=0}-\frac{\xi\beta p_3^2}{2\omega_i}f_i^{\xi=0}(1-f_i^{\xi=0})
~.\ee
The $\xi$-independent part of the quark distribution function in the 
presence of a strong magnetic field in general frame is written as
\be
f_i^{\xi=0}=\frac{1}{e^{\beta u^\alpha\tilde{p}_\alpha}+1}
~,\ee
where $\tilde{p}_\alpha\equiv\left(\omega_i,p_3\right)$ with 
$\omega_i$ in the strong magnetic field limit ($n=0$) is 
given by $\omega_i =\sqrt{p_3^2+m_i^2}$. 

The gluons which are electrically uncharged particles are no 
longer affected by the $B$-driven anisotropy. Thus the gluon 
distribution function retains its form as in the isotropic 
case. The quark contributions to the shear and bulk 
viscosities become modified due to the presence 
of anisotropy created by the strong magnetic field. In the SMF 
limit, only longitudinal (along the direction of magnetic field) 
shear and bulk viscosities have contributions from the lowest 
Landau level (LLL) quarks, so we are now going to calculate the 
longitudinal components of the viscosities.

In the presence of strong magnetic field, effective (1+1)-dimensional 
kinetic theory helps to determine transport coefficients. Due to 
dimensional reduction, the (integration) phase factor is written \cite{Gusynin:NPB462'1996,Bruckmann:PRD96'2017} as
\be\label{phase factor}
\int\frac{d^3{\rm p}}{(2\pi)^3}=\frac{|q_iB|}{2\pi}\int \frac{dp_3}{2\pi}
~.\ee
The energy-momentum tensor 
($\tilde{T}^{\mu\nu}=\tilde{T}_{(0)}^{\mu\nu}+\Delta\tilde{T}^{\mu\nu}$) in this 
regime has the following form,
\be
\tilde{T}^{\mu\nu}=\sum_i \frac{g_i|q_iB|}{2\pi^2}\int d p_3 ~ \frac{\tilde{p}^\mu\tilde{p}^\nu}{\omega_i}f_i
~.\ee
Similarly, the nonequilibrium part of the energy-momentum tensor is written as
\be\label{emb1}
\Delta\tilde{T}^{\mu\nu}=\sum_i \frac{g_i|q_iB|}{2\pi^2}\int d p_3 ~ \frac{\tilde{p}^\mu\tilde{p}^\nu}{\omega_i}\delta f_i
~,\ee
where the new notation for momentum $\tilde{p}^\mu$ in 
SMF limit is defined as $\tilde{p}^\mu=(p^0,0,0,p^3)$. 
The relativistic Boltzmann transport equation for 
quark distribution function in the relaxation-time approximation 
in conjunction with the strong magnetic field limit is written as
\begin{eqnarray}
\tilde{p}^\mu\partial_\mu f_i(x,p) = -\frac{\tilde{p}_\nu u^\nu}{\tau^B_i}\delta f_i
~.\end{eqnarray}
Here $\tau^B_i$ denotes the relaxation-time for quark in the 
presence of strong magnetic field and is given \cite{Hattori:PRD95'2017} by
\be
\tau^B_i=\frac{\omega_i\left(e^{\beta\omega_i}-1\right)}{\alpha_sC_2m_i^2\left(e^{\beta\omega_i}+1\right)}\left[1\Bigg{/}\left\lbrace\int dp^\prime_3\frac{1}{\omega^\prime_i\left(e^{\beta\omega^\prime_i}+1\right)}\right\rbrace\right]
,\ee
where $C_2$ is the Casimir factor. After substituting the value of 
$\delta f_i$ in eq. \eqref{emb1}, we get
\be\label{emb2}
\Delta\tilde{T}^{\mu\nu}=-\sum_i \frac{g_i|q_iB|}{2\pi^2}\int d p_3 ~ \frac{\tilde{p}^\mu\tilde{p}^\nu}{\tilde{p}_\nu u^\nu}\frac{\tau^B_i \tilde{p}^\mu\partial_\mu f_i}{\omega_i}
~.\ee

In the presence of weak-momentum anisotropy due to the strong 
magnetic field, the partial derivative of the anisotropic 
quark distribution function is calculated as
\be
\nonumber\partial_\mu f_{{\rm B},i}^{\rm aniso} &=& \frac{f_i^{\xi=0}(1-f_i^{\xi=0})}{T}\left[u_\alpha \tilde{p}^\alpha u_\mu\frac{DT}{T}+u_\alpha \tilde{p}^\alpha\frac{\nabla_\mu T}{T}-u_\mu \tilde{p}^\alpha Du_\alpha-\tilde{p}^\alpha\nabla_\mu u_\alpha\right] \\ && \nonumber -\frac{\xi p_3^2}{2}\left[-\frac{f_i^{\xi=0}(1-f_i^{\xi=0})}{\omega_i T^2}\left(u_\mu DT+\nabla_\mu T\right)\right. \\ && \left.-\frac{f_i^{\xi=0}(1-f_i^{\xi=0})}{\omega_i^2 T}\left(u_\mu \tilde{p}_\alpha Du^\alpha+\tilde{p}_\alpha\nabla_\mu u^\alpha\right)+\frac{1-2f_i^{\xi=0}}{\omega_i T}\partial_\mu f_i^{\xi=0}\right]
.\ee
Substituting the above expression in eq. \eqref{emb2} for the case 
of $B$-driven anisotropy and then calculating the space-space 
or longitudinal component of $\Delta\tilde{T}^{\mu\nu}$, we get
\be\label{emb3}
\nonumber\Delta\tilde{T}^{ij} &=& \sum_i \frac{g_i|q_iB|}{2\pi^2}\int d p_3 ~ \frac{\tilde{p}^i\tilde{p}^j}{\omega_i T} ~ \tau^B_i ~ f_i^{\xi=0}(1-f_i^{\xi=0})\left[\left\lbrace\omega_i\left(\frac{\partial P}{\partial\varepsilon}\right)-\frac{p_3^2}{3\omega_i}\right\rbrace{\partial}_lu^l-\frac{\tilde{p}^k\tilde{p}^l}{2\omega_i}W_{kl}\right. \\ && \left.\nonumber+\tilde{p}^k\left(\frac{{\partial}_k P}{\varepsilon+P}-\frac{{\partial}_k T}{T}\right)\right]-\xi\sum_i \frac{g_i|q_iB|}{2\pi^2}\int d p_3 ~ \frac{\tilde{p}^i\tilde{p}^jp_3^2}{2\omega_i^3 T} ~ \tau^B_i ~ f_i^{\xi=0}(1-f_i^{\xi=0}) \\ && \nonumber\times\left[-\left\lbrace\omega_i\left(\frac{\partial P}{\partial\varepsilon}\right)+\frac{p_3^2}{3\omega_i}\right\rbrace{\partial}_lu^l-\frac{\tilde{p}^k\tilde{p}^l}{2\omega_i}W_{kl}+\tilde{p}^k\left(\frac{{\partial}_k P}{\varepsilon+P}+\frac{{\partial}_k T}{T}\right)\right] \\ && \nonumber -\xi\sum_i \frac{g_i|q_iB|}{2\pi^2}\int d p_3 ~ \frac{\tilde{p}^i\tilde{p}^jp_3^2}{2\omega_i^2 T^2} ~ \tau^B_i ~ f_i^{\xi=0}(1-f_i^{\xi=0})(1-2f_i^{\xi=0}) \\ && \times\left[\left\lbrace\omega_i\left(\frac{\partial P}{\partial\varepsilon}\right)-\frac{p_3^2}{3\omega_i}\right\rbrace{\partial}_lu^l-\frac{\tilde{p}^k\tilde{p}^l}{2\omega_i}W_{kl}+\tilde{p}^k\left(\frac{{\partial}_k P}{\varepsilon+P}-\frac{{\partial}_k T}{T}\right)\right]
.\ee
The  pressure and the energy density in a strong magnetic field can be written 
in terms of the energy-momentum tensor as 
$P=-\Delta^\parallel_{\mu\nu}\tilde{T}^{\mu\nu}$ and 
$\varepsilon=u_\mu\tilde{T}^{\mu\nu}u_\nu$, respectively, where the 
longitudinal projection tensor is denoted by $\Delta^\parallel_{\mu\nu}=g^\parallel_{\mu\nu}-u_\mu u_\nu$ with  
$g^\parallel_{\mu\nu}$ (${\rm{diag}}(1,0,0,-1)$) as the 
suitable metric tensor. 

It is known that, instead of only two ordinary 
viscosity coefficients, $\eta$ and $\zeta$ (in eq. \eqref{definition})
in the absence of magnetic field, the eight coefficients suffice to
describe the viscous behavior in the presence of magnetic 
field, wherein the Onsager relation, however, reduces 
the numbers from eight to seven. The seven independent coefficients 
can be further grouped into the five shear viscosity coefficients - 
$\eta$, $\eta_1$, $\eta_2$, $\eta_3$ and $\eta_4$, one 
volume or bulk viscosity coefficient - $\zeta$ and a cross-effect 
between the ordinary and volume viscosities - $\zeta_1$. Thus, the 
linear combination of seven independent tensors yields the 
viscous tensor for an arbitrary magnetic field, ${\bf B}$ 
(with a direction, $\bf b =\frac{\bf B}{\rm B}$) \cite{Lifshitz:BOOK'1981}, 
\begin{eqnarray}\label{Form1}
\pi_{ij} &=& 2 \eta\left(V_{ij}-\frac{1}{3}
\delta_{ij} \nabla \cdot \mathbf V \right) +\zeta 
\delta_{ij} \nabla \cdot \mathbf V\nonumber \\
&& +\eta_1\left(2V_{ij}-\delta_{ij}\nabla\cdot\mathbf{V} +
\delta_{ij}V_{kl}b_k b_l-2V_{ik}
b_k b_j-2V_{jk}b_k b_i+b_i b_j\nabla
\cdot\mathbf{V}+b_i b_j V_{kl}b_k b_l\right) 
\nonumber \\
&& +2\eta_2\left(V_{ik}b_k b_j+V_{jk}b_k b_i-2b_i b_j V_{kl}b_k b_l
\right)\nonumber \\
&& +\eta_3\left(V_{ik}b_{jk}+V_{jk}
b_{ik}-V_{kl}b_{ik}b_j b_l-
V_{kl}b_{jk}b_i b_l\right) \nonumber \\
&& +2\eta_4\left(V_{kl}b_{ik}b_j b_l +
V_{kl}b_{jk}b_i b_l\right)\nonumber \\
&& +\zeta_1\left(\delta_{ij}V_{kl}b_k b_l+
b_i b_j\nabla\cdot\mathbf{V}\right)
,\end{eqnarray}
which is broadly decomposed into the traceless components (which are the 
coefficients of $\eta$, $\eta_1$, $\eta_2$, $\eta_3$, $\eta_4$) and the 
nonzero traces (the coefficients of $\zeta$ and $\zeta_1$).
The usual symbols used in the above equation are
\begin{eqnarray*}
&&b_{ij}=\epsilon_{ijk}b_k, \\ 
&&V_{ij}=\frac{1}{2}
\left(\frac{\partial V_i}{\partial x_j}
+\frac{\partial V_j}{\partial x_i}\right)
.\end{eqnarray*}
The first two terms in eq. \eqref{Form1} are the 
usual terms at B=0, so $\eta$ and $\zeta$ are the ordinary viscosity 
coefficients. 

When applied to plasma, the above tensor \eqref{Form1} is simplified 
by the vanishing of the cross effect between ordinary 
viscosity and volume viscosity ($\zeta_1$). The tensor 
could be further reduced in a much simpler form 
in the strong magnetic field by the vanishing 
of $\eta_1,\eta_2, \eta_3$ and $\eta_4$ coefficients. This can be easily seen
by first replacing the $\eta$-term in the tensor,
\[
\eta_0 \left(3 b_i b_j - \delta_{ij}\right) \left(b_k b_l V_{kl} -\frac{1}{3} \nabla \cdot V\right),
\]
and then rearranging the terms in the tensor. Thus, the components 
of the tensor \eqref{Form1} in a magnetic field along a specific 
direction ($z$-direction) are written in Cartesian coordinates as
\begin{eqnarray}
&&\pi_{xx} = -\eta_0\left(V_{zz}-\frac{1}{3}\nabla\cdot
\mathbf{V}\right)+\eta_1\left(V_{xx}-V_{yy}\right)+2\eta_3 V_{xy}+
\zeta_0\nabla\cdot\mathbf{V} , \\
&&\pi_{yy} = -\eta_0\left(V_{zz}-\frac{1}{3}\nabla\cdot
\mathbf{V}\right)+\eta_1\left(V_{yy}-V_{xx}\right)-2\eta_3V_{xy}+
\zeta_0\nabla\cdot\mathbf{V}, \\
&&\pi_{zz} = 2\eta_0\left(V_{zz}-\frac{1}{3}\nabla\cdot
\mathbf{V}\right)+\zeta_0 \nabla\cdot\mathbf{V}, \\
&&\pi_{xy} = 2\eta_1V_{xy}-\eta_3\left(V_{xx}-V_{yy}\right), \\
&&\pi_{xz} = 2\eta_2 V_{xz}+2\eta_4 V_{yz}, \\
&&\pi_{yz} = 2\eta_2 V_{yz} - 2\eta_4 V_{xz}
.\end{eqnarray}
When the magnetic field becomes strong, the motion is 
restricted to one-dimension in the direction of the magnetic field. 
As a result, the transverse components of the velocity
gradient - $V_{xx},V_{yy},V_{xy}$ vanish, which in turn make
the non-diagonal terms of the tensor - $\pi_{xy}$, $\pi_{xz}$ 
and $\pi_{yz}$ zero. Thus, the nonvanishing (longitudinal) components in 
the viscous tensor are written as 
\begin{eqnarray}
&&\pi_{xx} = -\eta_0\left(V_{zz}-\frac{1}{3}\nabla\cdot
\mathbf{V}\right)+\zeta_0\nabla\cdot\mathbf{V} , \\
&&\pi_{yy} = -\eta_0\left(V_{zz}-\frac{1}{3}\nabla\cdot
\mathbf{V}\right)+\zeta_0\nabla\cdot\mathbf{V}, \\
&&\pi_{zz} = 2\eta_0\left(V_{zz}-\frac{1}{3}\nabla\cdot
\mathbf{V}\right)+\zeta_0 \nabla\cdot\mathbf{V}
,\end{eqnarray}
where $\eta_0$ and $\zeta_0$ are known as the longitudinal 
viscosities.\footnote{The term longitudinal signifies the direction 
of the velocity with respect to the direction of magnetic field.} 

The above components consist of traceless and nonzero trace terms 
and the coefficients of them are the shear and bulk viscosities, 
respectively, {\em like} the case in the absence of magnetic field 
in eq. \eqref{definition}. Hence, separating the traceless and nonzero trace 
parts, the above 
components are grouped into forms, 
\begin{eqnarray}
&&\pi_{xx}=\pi_{yy}=-\frac{1}{2}\pi_{zz}=-\eta_0
\left(V_{zz}-\frac{1}{3} {\nabla\cdot \mathbf{V}|}_z \right), \\ 
&&\pi_{xx}=\pi_{yy}=\pi_{zz}=\zeta_0 {\nabla\cdot\mathbf{V}|}_z
,\end{eqnarray}
respectively. The coefficients of those traceless and nonzero trace terms
are the (longitudinal) shear and bulk viscosities, respectively. 
Therefore, generalizing the viscous tensor into the 
relativistic energy-momentum tensor, ${\tilde{T}}^{\mu \nu}$ 
\cite{Lifshitz:BOOK'1981,Ofengeim:EPL112'2015} in the strong magnetic field 
regime, the spatial component of the dissipative part of the relativistic 
energy-momentum tensor can be defined (appendix \ref{form.tensor}) as
(by relabeling $\eta_0 \equiv \eta^B$ and $\zeta_0 \equiv \zeta^B$
as an artifact of the strong magnetic field limit),
\be\label{definition(eb)}
\Delta\tilde{T}^{ij}=-\eta^B W^{ij}-\zeta^B\delta^{ij}{\partial}_l u^l
.\ee

From equations \eqref{emb3} and \eqref{definition(eb)}, we get the 
quark contribution to the shear viscosity for the $B$-driven 
anisotropic medium as
\be\label{A.S.V.q(eB)}
\nonumber\eta_{\rm B,q}^{\rm aniso} &=& \frac{\beta}{4\pi^2}\sum_i g_i ~ |q_iB|\int dp_3~\frac{p_3^4}{\omega_i^2} ~ \tau_i^B ~ f_i^{\xi=0}(1-f_i^{\xi=0}) \\ && - \frac{\xi\beta^2}{8\pi^2}\sum_i g_i ~ |q_iB|\int d{p_3}~\frac{p_3^6}{\omega_i^3}~ \tau_i^B ~ f_i^{\xi=0}(1-f_i^{\xi=0})\left\lbrace \frac{1}{\beta\omega_i}+1-2f_i^{\xi=0} \right\rbrace
.\ee
Since gluons are not influenced by the presence of magnetic field, the 
gluon part of the shear viscosity remains unaffected by the $B$-driven 
anisotropy. So we can add the isotropic gluon contribution to obtain the total 
shear viscosity,
\be\label{A.S.V.(eB)}
\nonumber\eta_{\rm B}^{\rm aniso} &=& \frac{\beta}{4\pi^2}\sum_i g_i ~ |q_iB|\int dp_3~\frac{p_3^4}{\omega_i^2} ~ \tau_i^B ~ f_i^{\xi=0}(1-f_i^{\xi=0}) \\ && - \nonumber\frac{\xi\beta^2}{8\pi^2}\sum_i g_i ~ |q_iB|\int d{p_3}~\frac{p_3^6}{\omega_i^3}~ \tau_i^B ~ f_i^{\xi=0}(1-f_i^{\xi=0})\left\lbrace \frac{1}{\beta\omega_i}+1-2f_i^{\xi=0} \right\rbrace \\ && +\frac{\beta}{30\pi^2} g_g \int d{\rm p}~\frac{{\rm p}^6}{\omega_g^2} ~ \tau_g ~ f_g^{\rm iso}(1+f_g^{\rm iso})
~,\ee
which can further be decomposed as
\be\label{A.S.V.(1eB)}
\nonumber\eta_{\rm B}^{\rm aniso} &=& \eta^{\xi=0}+\eta^{\xi\ne 0} \\ &=& \eta^{\xi=0} - \frac{\xi\beta^2}{8\pi^2}\sum_i g_i ~ |q_iB|\int d{p_3}~\frac{p_3^6}{\omega_i^3}~ \tau_i^B ~ f_i^{\xi=0}(1-f_i^{\xi=0})\left\lbrace \frac{1}{\beta\omega_i}+1-2f_i^{\xi=0} \right\rbrace
.\ee
The bulk viscosity due to quark contribution can also be obtained by 
comparing equations \eqref{emb3} and \eqref{definition(eb)},
\be\label{A.B.V.q(eB)}
\nonumber\zeta_{\rm B,q}^{\rm aniso} &=& \sum_i \frac{g_i|q_iB|}{2\pi^2}\int d p_3~\frac{p_3^2}{\omega_i} ~ f_i^{\xi=0}(1-f_i^{\xi=0})A_{1,i} \\ && - \nonumber\xi\sum_i \frac{g_i|q_iB|}{2\pi^2}\int d p_3~\frac{p_3^4}{2\omega_i^3} ~ f_i^{\xi=0}(1-f_i^{\xi=0})A_{2,i} \\ && -\xi\sum_i \frac{g_i|q_iB|}{2\pi^2}\int d p_3~\frac{p_3^4}{2\omega_i^2 T} ~ f_i^{\xi=0}(1-f_i^{\xi=0})(1-2f_i^{\xi=0})A_{1,i}
~,\ee
where $A_{1,i}$ and $A_{2,i}$ have the following forms,
\begin{eqnarray}
&&A_{1,i}=\frac{\tau_i^B}{3T}\left[\frac{p_3^2}{\omega_i}-3\left(\frac{\partial P}{\partial \varepsilon}\right)\omega_i\right],\label{A1i} \\ 
&&A_{2,i}=\frac{\tau_i^B}{3T}\left[\frac{p_3^2}{\omega_i}+3\left(\frac{\partial P}{\partial \varepsilon}\right)\omega_i\right]\label{A2i}
.\end{eqnarray}
Applying the Landau-Lifshitz condition for the calculation 
of the bulk viscosity and then simplifying, we get
\be\label{A.B.V.q1(eB)}
\nonumber\zeta_{\rm B,q}^{\rm aniso} &=& \frac{\beta}{6\pi^2}\sum_i g_i|q_iB|\int d p_3~\left[\frac{p_3^2}{\omega_i}-3\left(\frac{\partial P}{\partial\varepsilon}\right)\omega_i\right]^2 \tau_i^B f_i^{\xi=0}(1-f_i^{\xi=0}) \\ && - \nonumber\frac{\xi\beta}{12\pi^2}\sum_i g_i|q_iB|\int d p_3~\frac{p_3^2}{\omega_i^2}\left[\frac{p_3^4}{\omega_i^2}-9\left(\frac{\partial P}{\partial\varepsilon}\right)^2\omega_i^2\right] \tau_i^B f_i^{\xi=0}(1-f_i^{\xi=0}) \\ && - \nonumber\frac{\xi\beta^2}{12\pi^2}\sum_i g_i|q_iB|\int d p_3~\frac{p_3^2}{\omega_i}\left[\frac{p_3^2}{\omega_i}-3\left(\frac{\partial P}{\partial\varepsilon}\right)\omega_i\right]^2 \tau_i^B f_i^{\xi=0}(1-f_i^{\xi=0}) \\ && \hspace{9.3 cm}\times(1-2f_i^{\xi=0})
~.\ee
As has been mentioned earlier that the $B$-driven anisotropy has no 
influence on gluons, so the total bulk viscosity can be obtained by 
adding the isotropic gluon contribution to the modified quark 
contribution as follows,
\be\label{A.B.V.(eB)}
\nonumber\zeta_{\rm B}^{\rm aniso} &=& \frac{\beta}{6\pi^2}\sum_i g_i|q_iB|\int d p_3~\left[\frac{p_3^2}{\omega_i}-3\left(\frac{\partial P}{\partial\varepsilon}\right)\omega_i\right]^2 \tau_i^B f_i^{\xi=0}(1-f_i^{\xi=0}) \\ && - \nonumber\frac{\xi\beta}{12\pi^2}\sum_i g_i|q_iB|\int d p_3~\frac{p_3^2}{\omega_i^2}\left[\frac{p_3^4}{\omega_i^2}-9\left(\frac{\partial P}{\partial\varepsilon}\right)^2\omega_i^2\right] \tau_i^B f_i^{\xi=0}(1-f_i^{\xi=0}) \\ && - \nonumber\frac{\xi\beta^2}{12\pi^2}\sum_i g_i|q_iB|\int d p_3~\frac{p_3^2}{\omega_i}\left[\frac{p_3^2}{\omega_i}-3\left(\frac{\partial P}{\partial\varepsilon}\right)\omega_i\right]^2 \tau_i^B f_i^{\xi=0}(1-f_i^{\xi=0})(1-2f_i^{\xi=0}) \\ && +\frac{\beta}{18\pi^2}g_g\int d{\rm p}~{\rm p}^2\left[\frac{{\rm p}^2}{\omega_g}-3\left(\frac{\partial P}{\partial \varepsilon}\right)\omega_g\right]^2 \tau_g ~ f_g^{\rm iso}(1+f_g^{\rm iso})
~,\ee
which can be written in terms of $\xi$-independent and 
$\xi$-dependent parts as
\be\label{A.B.V.(1eB)}
\nonumber\zeta_{\rm B}^{\rm aniso} &=& \zeta^{\xi=0}+\zeta^{\xi\neq 0} \\ &=& \zeta^{\xi=0} - \nonumber\xi\left[\frac{\beta^2}{12\pi^2}\sum_i g_i|q_iB|\int d p_3~\frac{p_3^2}{\omega_i} ~ \tau_i^B f_i^{\xi=0}(1-f_i^{\xi=0})\right. \\ && \left.\times\left\lbrace\frac{1}{\beta\omega_i}\left[\frac{p_3^4}{\omega_i^2}-9\left(\frac{\partial P}{\partial\varepsilon}\right)^2\omega_i^2\right]+(1-2f_i^{\xi=0})\left[\frac{p_3^2}{\omega_i}-3\left(\frac{\partial P}{\partial\varepsilon}\right)\omega_i\right]^2\right\rbrace\right]
.\ee
\begin{figure}[]
\begin{center}
\begin{tabular}{c c}
\includegraphics[width=7cm]{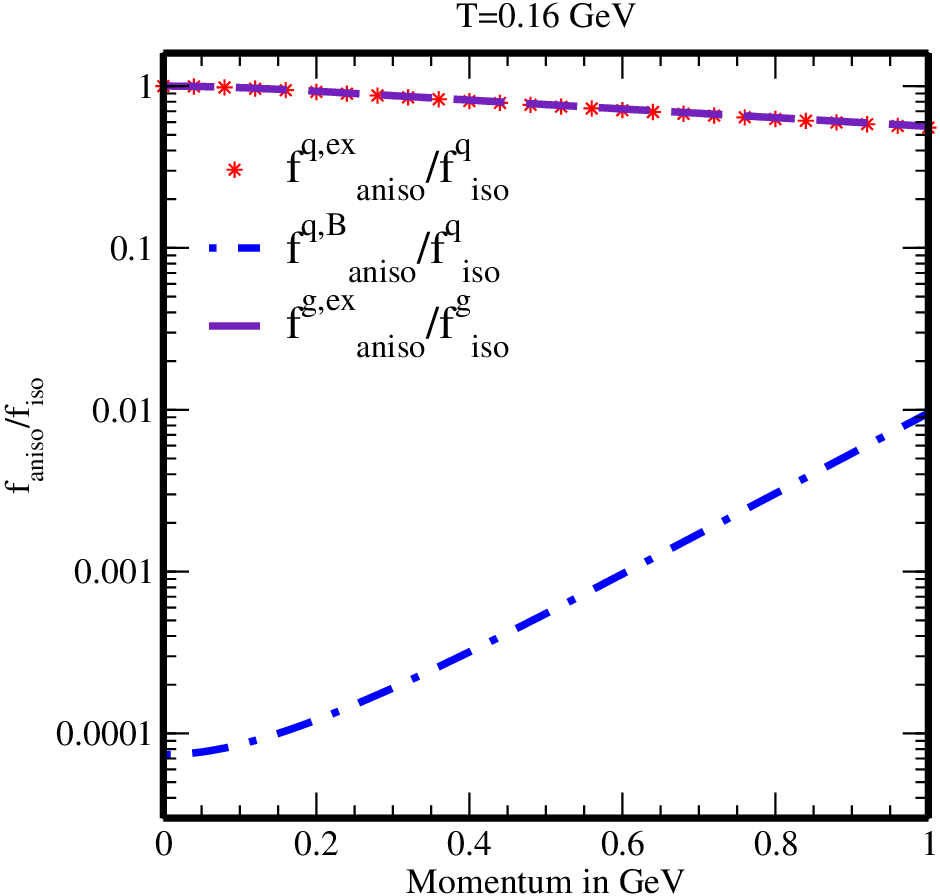}&
\hspace{0.423 cm}
\includegraphics[width=7cm]{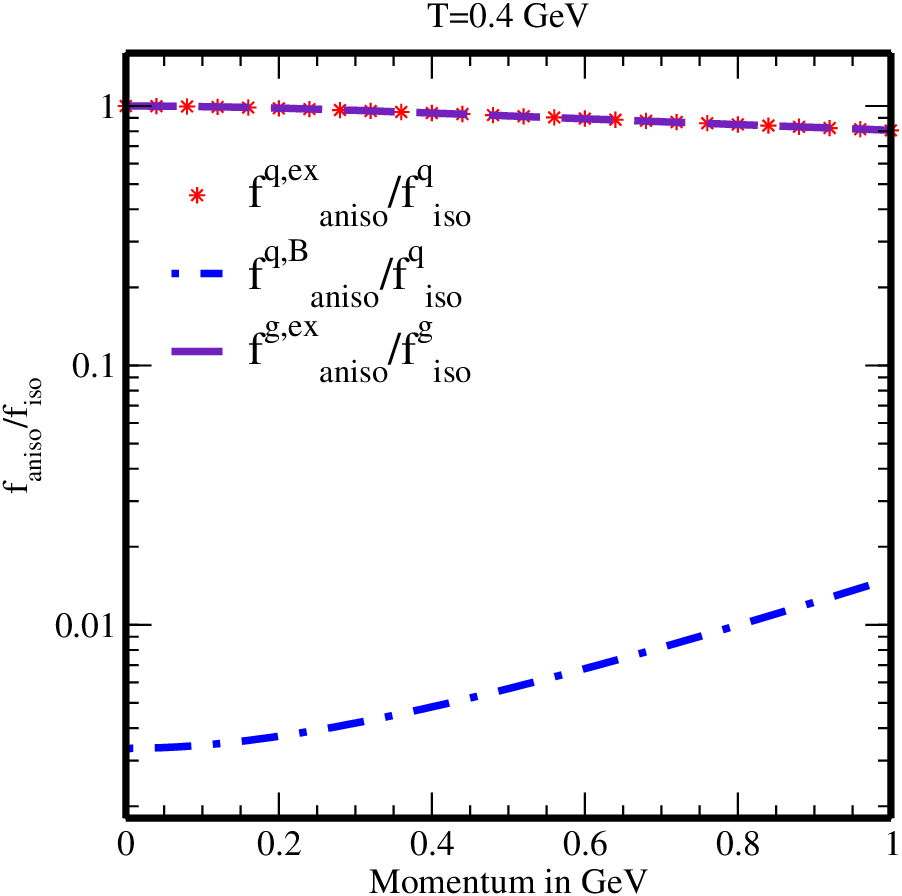} \\
a & b
\end{tabular}
\caption{Variation of the ratio $f_{\rm aniso}/f_{\rm iso}$ with momentum 
in the presence of momentum anisotropies both due to asymptotic 
expansion and strong magnetic field ($15$ $m_\pi^2$) at (a) low temperature 
and (b) high temperature with the quasiparticle masses for quarks and
gluons.}\label{fup.qpm}
\end{center}
\end{figure}

\begin{figure}[]
\begin{center}
\begin{tabular}{c c}
\includegraphics[width=7cm]{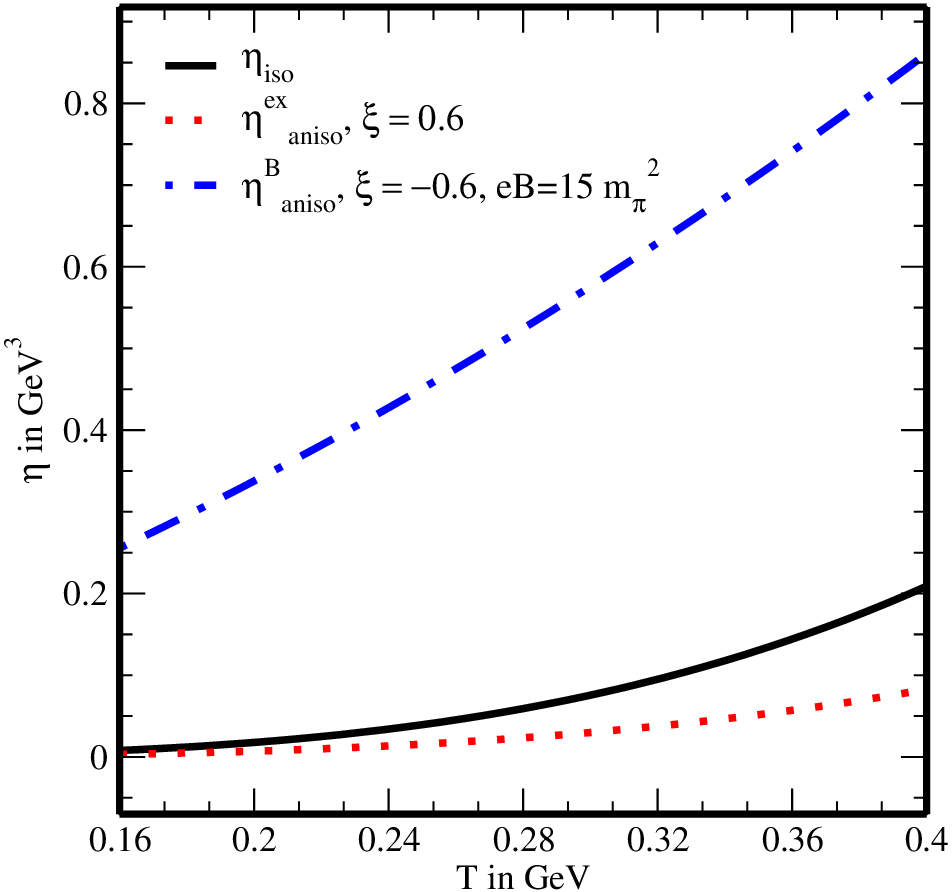}&
\hspace{0.423 cm}
\includegraphics[width=7cm]{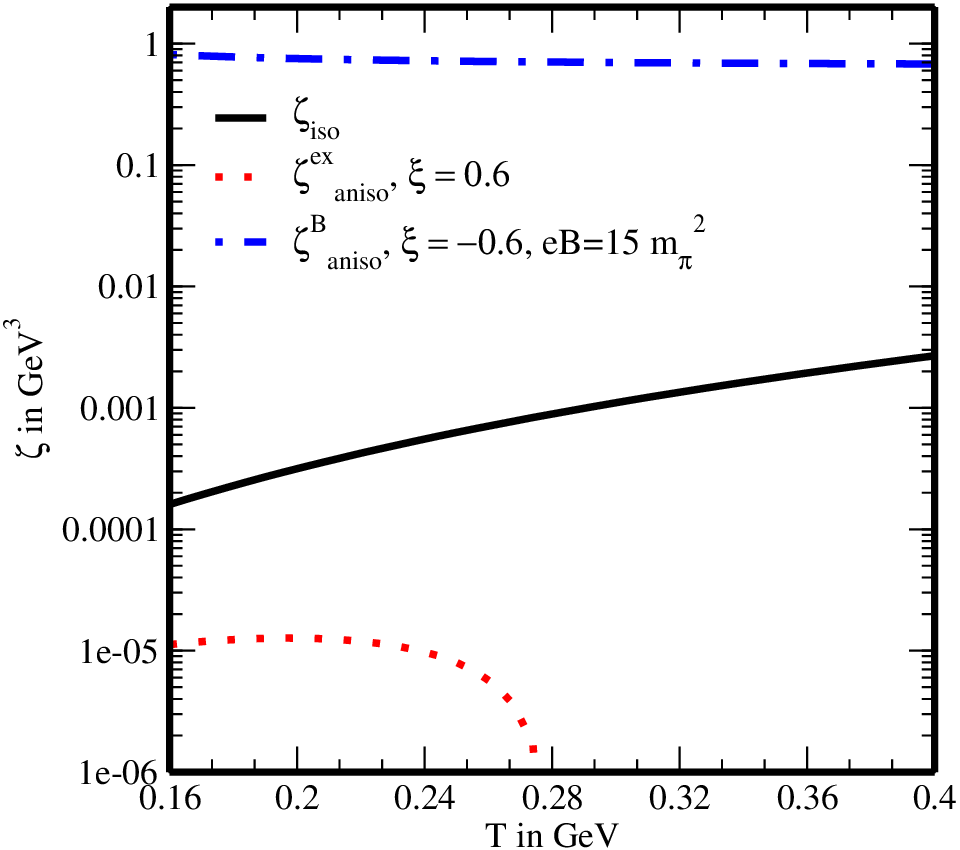} \\
a & b
\end{tabular}
\caption{Variations of (a) the shear viscosity and (b) the bulk viscosity 
with temperature in the presence of momentum anisotropies both due to 
asymptotic expansion and strong magnetic field.}\label{sb.1}
\end{center}
\end{figure}

Before discussing the results on the shear viscosity and bulk 
viscosity in the presence of magnetic field-induced and 
expansion-induced anisotropies, it is utmost important to 
understand the behaviors of the isotropic and anisotropic 
distribution functions, because the 
behaviors of transport coefficients mainly depend on the phase-space 
factor, relaxation-time and the distribution function which in general 
embraces all the 
information on the influence of anisotropy. Thus, it becomes essential to 
explore the effects of anisotropies on quark and gluon distribution
functions through their ratios with respect to their isotropic 
counterparts, {\em viz.} $\dex/\df$, $\db/\df$, $\dgex/\dg$ 
in figure \ref{fup.qpm} at two temperatures. We have employed the 
quasiparticle description in the distribution functions for the isotropic 
and expansion-driven anisotropic mediums by the $T$-dependent masses 
for gluons \eqref{Gluon mass} and quarks \eqref{Quark mass}, 
whereas the $T$ and $B$-dependent mass \eqref{Mass} has been used
in the distribution function for the $B$-driven anisotropic 
medium. It is found that the 
effects of anisotropy caused by the expansion on quark and gluon 
distributions are almost identical (seen in figure \ref{fup.qpm}), at 
least for the weak-anisotropic limit. However, the ratios get decreased 
in the high momentum regime. In the presence of strong magnetic field the 
distribution function for quark gets affected severely and the ratio in 
low momenta is tiny and increases at higher momenta. 
With the aforesaid findings on the distribution functions in the
presence of anisotropies, we have computed the shear viscosity 
in isotropic \eqref{iso.eta}, expansion- \eqref{A.S.V.(1)} and 
$B$-driven anisotropic \eqref{A.S.V.(1eB)} mediums and the 
bulk viscosity in isotropic \eqref{iso.zeta}, expansion- \eqref{A.B.V.(1)} 
and $B$-driven anisotropic \eqref{A.B.V.(1eB)} mediums. 
From figure \ref{sb.1}a we have observed that, at low temperatures, the 
difference between the values of $\eta$ in isotropic medium and in the 
presence of weak-momentum anisotropy ($\xi=0.6$) due to 
asymptotic expansion is almost negligible, however, 
with the increase of temperature, this difference gradually 
increases, {\em i.e.} $\eta$ becomes smaller than its isotropic 
counterpart. If the origin of weak-momentum anisotropy is strong magnetic 
field, then the magnitude of $\eta$ becomes higher than that 
in isotropic medium and with temperature, this difference increases. Thus 
the above anisotropies leave different imprints on the shear viscosity, 
which are attributed mainly by the modified distribution function, phase
space factor and relaxation-time in the absence and presence of
strong magnetic field. Similarly, $\zeta$ gets amplified in $B$-driven 
anisotropy compared to both isotropic and expansion-driven anisotropic cases 
(in figure \ref{sb.1}b). However with the increase of temperature, 
$\zeta$ decreases very slowly, opposite to a slow increase in isotropic 
medium. Interestingly, if the anisotropy is originated from the initial 
asymptotic expansion, then $\zeta$ becomes meagre and approaches 
zero at a higher temperature. 

\subsection{Ratios of the shear ($\eta/s$) and bulk ($\zeta/s$) viscosities to the entropy density}
\begin{figure}[]
\begin{center}
\includegraphics[width=7cm]{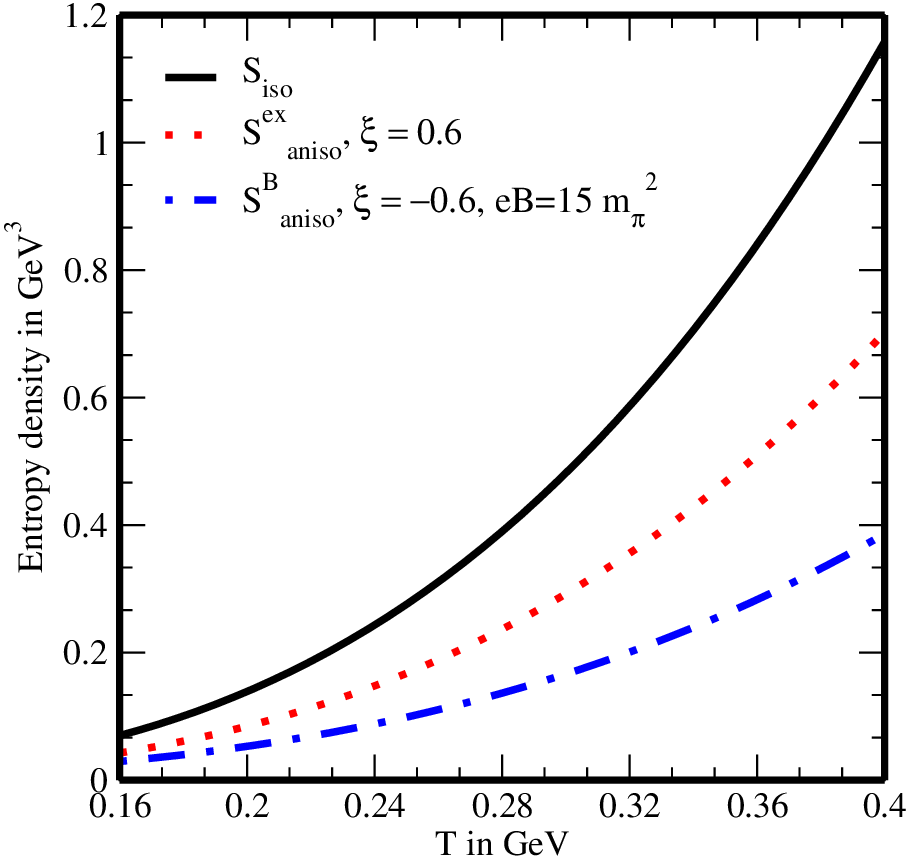}
\caption{Variation of the entropy density 
with temperature in the presence of momentum anisotropies 
both due to asymptotic expansion and strong 
magnetic field.}\label{ed.1}
\end{center}
\end{figure}

We are now going to study the effects of momentum anisotropies generated at the
early stages of collisions in URHICs on the dimensionless ratios, 
$\eta/s$ and $\zeta/s$, because they are useful
in characterizing how close the matter produced at URHICs 
is to being perfect and conformal fluid, respectively. The phenomenological 
studies by parton transport of the collective behavior \cite{Xu:PRL101'2008,Ferini:PLB670'2009,Cassing:NPA831'2009,Bratkovskaya:NPA856'2011} have reported that the QGP 
has a very small value of $\eta/s \approx \frac{1}{4\pi}$, suggesting that
the matter produced at RHIC is a strongly-coupled fluid of quarks
and gluons, contrary to the belief of weakly interacting gas of quarks and 
gluons on the basis of asymptotic freedom. Similarly, the study of AdS/CFT 
correspondence \cite{Kovtun:PRL94'2005} constrains the value of $\eta/s$ 
by a lower bound of $\frac{1}{4\pi}$. The hydrodynamic 
model~\cite{Luzum:PRC78'2008} also with small value of $\eta/s$ ranging 
from $\frac{1}{4\pi}$ to $\frac{2}{4\pi}$ consistently reproduces 
the experimental data 
\cite{Gavin:PRL97'2006,Drescher:PRC76'2007} and lattice calculations 
\cite{Nakamura:PRL94'2005,Meyer:PRD76'2007}. The bulk viscosity 
is yet to be developed at the early times of the hydrodynamic evolution, 
so some early viscous hydrodynamic simulations have usually 
ignored it in the dissipative part of energy-momentum tensor 
for simplicity \cite{Muronga:PRL88'2002,Song:JPG36'2009}. 
Although $\zeta$ vanishes for a thermal QCD medium of massless flavors 
on the classical level due to the conformal symmetry, but the nonabelian 
interactions break the conformal symmetry of QCD and generate a nonzero 
bulk viscosity, which is found in the lattice calculation of SU(3) 
gauge theory \cite{Meyer:PRL100'2008}. Near the critical or crossover temperature 
of hadron to QGP phase transition, the value of $\zeta/s$ becomes a 
maximum whereas that of $\eta/s$ becomes a minimum. Thus, it becomes 
worthwhile to observe the behaviors of both $\eta/s$ and $\zeta/s$ in the 
presence of $B$- and expansion-induced anisotropies, which in turn gives the 
effect of strong magnetic field through the anisotropy it generated. 
In order to do this, one thus requires the expression of the entropy 
density ($s$) in the presence of anisotropies, which could be best 
derived in the abovementioned kinetic theory approach. For the 
chemical potential of quarks, $\mu_q=0$, the entropy density is obtained 
from the energy density and pressure by the relation,
\begin{eqnarray}\label{E.D.}
S=\frac{\varepsilon+P}{T}
~.\end{eqnarray}
Therefore we have first calculated the energy density and pressure
in isotropic as well as in anisotropic mediums in appendix \ref{A.(e,p,s)}, 
using the kinetic theory. Hence the above relation \eqref{E.D.} has been 
used to obtain the entropy densities for isotropic, expansion-driven anisotropic 
and $B$-driven anisotropic mediums as
\begin{eqnarray}
S^{\rm iso} &=& \frac{\beta}{3\pi^2}\sum_i g_i \int d{\rm p}~{\rm p}^2\left(\frac{{\rm p}^2}{\omega_i}+3\omega_i\right)f_i^{\rm iso}+\frac{\beta}{6\pi^2}g_g\int d{\rm p}~{\rm p}^2\left(\frac{{\rm p}^2}{\omega_g}+3\omega_g\right)f_g^{\rm iso}
, \label{iso.(E.D.)} \\ 
\nonumber S_{\rm ex}^{\rm aniso} &=& S^{\rm iso}-\frac{\xi\beta^2}{18\pi^2}\sum_i g_i \int d{\rm p}~\frac{{\rm p}^4}{\omega_i}\left(\frac{{\rm p}^2}{\omega_i}+3\omega_i\right)f_i^{\rm iso}(1-f_i^{\rm iso}) \\ && \nonumber-\frac{\xi\beta^2}{36\pi^2}g_g\int d{\rm p}~\frac{{\rm p}^4}{\omega_g}\left(\frac{{\rm p}^2}{\omega_g}+3\omega_g\right)f_g^{\rm iso}(1+f_g^{\rm iso}) \\ &=& \nonumber S^{\rm iso}-\xi\left[\frac{\beta^2}{18\pi^2}\sum_i g_i \int d{\rm p}~\frac{{\rm p}^4}{\omega_i}\left(\frac{{\rm p}^2}{\omega_i}+3\omega_i\right)f_i^{\rm iso}(1-f_i^{\rm iso})\right. \\ && \left.+\frac{\beta^2}{36\pi^2}g_g\int d{\rm p}~\frac{{\rm p}^4}{\omega_g}\left(\frac{{\rm p}^2}{\omega_g}+3\omega_g\right)f_g^{\rm iso}(1+f_g^{\rm iso})\right]
, \label{exaniso.(E.D.)} \\ 
\nonumber S_{\rm B}^{\rm aniso} &=& \frac{\beta}{2\pi^2}\sum_i g_i|q_iB|\int d p_3\left(\frac{p_3^2}{\omega_i}+\omega_i\right)f_i^{\xi=0} \\ && \nonumber-\frac{\xi\beta^2}{4\pi^2}\sum_i g_i|q_iB|\int d p_3 ~ \frac{p_3^2}{\omega_i}\left(\frac{p_3^2}{\omega_i}+\omega_i\right)f_i^{\xi=0}(1-f_i^{\xi=0}) \\ && \nonumber+\frac{\beta}{6\pi^2}g_g\int d{\rm p}~{\rm p}^2\left(\frac{{\rm p}^2}{\omega_g}+3\omega_g\right)f_g^{\rm iso} \\ &=& S^{\xi=0}-\frac{\xi\beta^2}{4\pi^2}\sum_i g_i|q_iB|\int d p_3 ~ \frac{p_3^2}{\omega_i}\left(\frac{p_3^2}{\omega_i}+\omega_i\right)f_i^{\xi=0}(1-f_i^{\xi=0}) \label{baniso.(E.D.)}
,\end{eqnarray}
respectively. The immediate observation is that the entropy 
density gets decreased in the presence of momentum anisotropy 
(seen in figure \ref{ed.1}), especially it is lowest in 
$B$-driven anisotropy due to the severe reduction of 
phase-space in the presence of strong magnetic field.

\begin{figure}[]
\begin{center}
\begin{tabular}{c c}
\includegraphics[width=7cm]{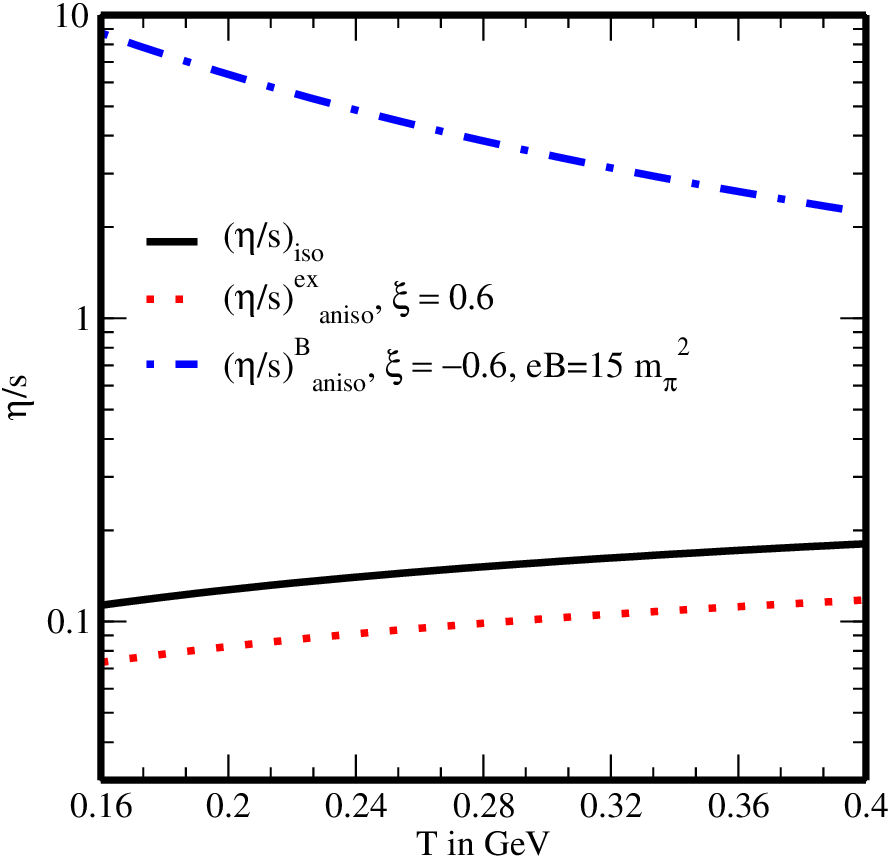}&
\hspace{0.423 cm}
\includegraphics[width=7cm]{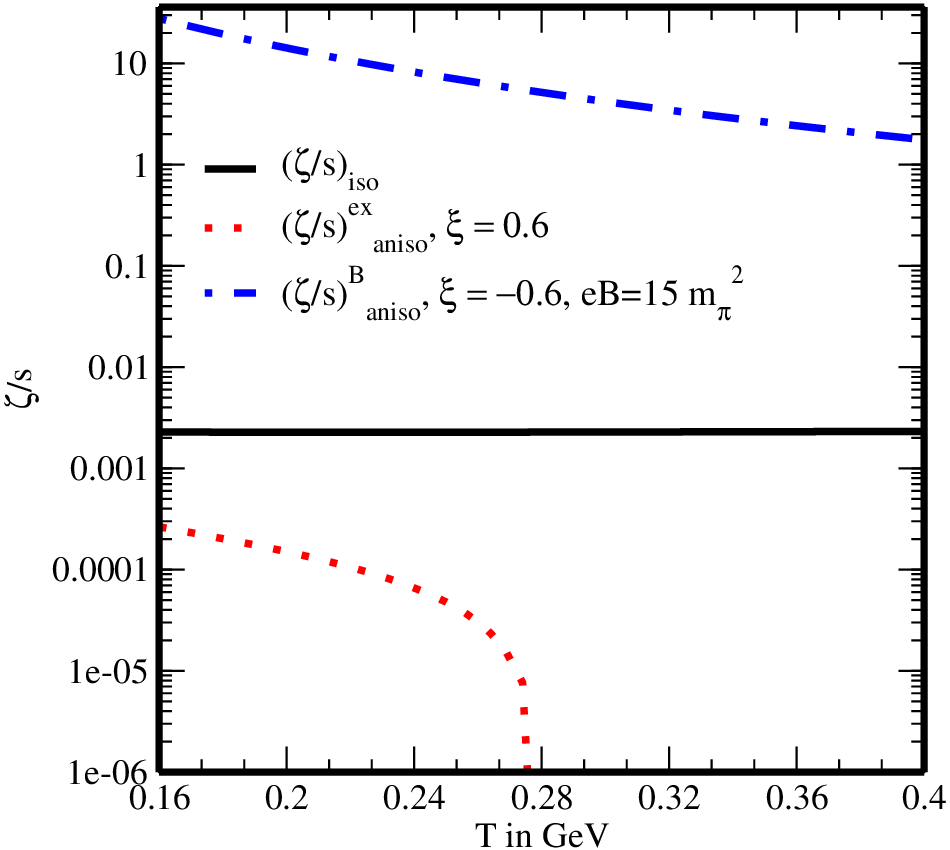} \\
a & b
\end{tabular}
\caption{Variations of (a) ${\eta}/{s}$ and (b) ${\zeta}/{s}$
with temperature in the presence of momentum anisotropies 
both due to asymptotic expansion and strong 
magnetic field.}\label{ratios.1}
\end{center}
\end{figure}

Thus, having the knowledge of entropy density in the presence 
of anisotropies, we have visualized the effects of anisotropies 
on the variations of $\eta/s$ and $\zeta/s$ with temperature in 
figures \ref{ratios.1}a and \ref{ratios.1}b, respectively. 
Since $s$ is always smaller than $\eta$ 
in $B$-driven anisotropy, $\eta/s$ is always larger than one, but 
unlike $\eta$ (as well as $s$), $\eta/s$ decreases with temperature 
(dashed-dotted line in figure \ref{ratios.1}a) because entropy 
density increases faster with $T$ than $\eta$. 
On the other hand, $\eta/s$ becomes much smaller ($<1$) in isotropic 
medium as well as in expansion-driven anisotropic medium (denoted by 
solid and dotted lines, respectively in figure \ref{ratios.1}a) than that 
in $B$-driven anisotropy, but $\eta/s$ increases with temperature monotonically, 
resulting finally the inequality: $\frac{\eta}{s}\big{|}_{\rm B-driven~aniso}>
\frac{\eta}{s}\big{|}_{\rm iso}>
\frac{\eta}{s}\big{|}_{\rm ex-driven~aniso}$. Shear viscosity in the 
isotropic case is known as collisional viscosity and the same arising 
due to weak-momentum anisotropy is called anomalous viscosity. 
In the theory of particle transport in turbulent plasma \cite{Asakawa:PRL96'2006}, 
it has been argued that, due to anomalous viscosity, even a 
weakly-coupled but expanding quark-gluon plasma may 
gain the character of a nearly perfect fluid, thus a large anisotropy 
describes a small value of anomalous viscosity. 
In our finding, the collisional viscosity comes out higher than the 
anomalous viscosity in expansion-driven anisotropy, thus the ratio 
$\eta/s$ indicates the character of nearly perfect 
fluid. On the other hand, the collisional viscosity is smaller than the 
anomalous viscosity in $B$-driven anisotropy, so $\eta/s$ takes the medium 
slightly away from the fluid character. 
Last but not the least, $\zeta/s$ is very small compared to $\eta/s$ 
except that in $B$-driven anisotropy, where it becomes comparable to 
$\eta/s$ and decreases with temperature (in figure \ref{ratios.1}b). 
However, like the variation of $\zeta$ with temperature, $\zeta/s$ in 
expansion-driven anisotropy vanishes at some higher temperature, which 
could have a resemblance with the temperature where the chiral symmetry is restored. 

\section{The coefficients affiliated to momentum, heat and charge transports}
In this section, we are going to study the effects of anisotropies 
on the relative behaviors among momentum, heat and charge transports 
through the Prandtl number, the Reynolds number and the ratio between 
momentum diffusion and charge diffusion. To be specific, the $B$-driven 
anisotropy in a way reveals the effect of strong magnetic field on the 
abovementioned transport coefficients. 

\subsection{Prandtl number}
The heat transfer and the momentum transfer in a medium are diffusive 
processes. The relative behavior between the momentum diffusion 
and the thermal diffusion can be described in terms of the Prandtl number, 
\begin{equation}\label{Pl}
{\rm Pl}=\frac{\eta/\rho}{\kappa/C_p}
~,\end{equation}
where $C_p$ is the specific heat at constant pressure, $\rho$ 
denotes the mass density and $\kappa$ represents the thermal 
conductivity. Thus, Pl describes the roles of thermal conductivity 
and shear viscosity on the sound attenuation in the system and has 
been calculated in a varieties of 
systems, {\em such as}, strongly coupled liquid helium \cite{T:RPP72'2009}, 
nonrelativistic conformal holographic fluid 
\cite{T:RPP72'2009,Rangamani:JHEP01'2009} and a dilute atomic 
Fermi gas \cite{Braby:PRA82'2010} etc. The Prandtl number sheds light 
on the sound attenuation in the system, which in turn tells about the 
energy loss while sound propagates in a 
medium. The Prandtl number of magnitude less than one implies the dominance 
of thermal diffusion over momentum diffusion in the sound attenuation, 
whereas the opposite happens for Pl greater than one. In this work, 
we wish to find out how the presence of momentum anisotropies 
in a medium could affect the competition between momentum and 
heat diffusions, resulting in the energy dissipation of sound propagation. 
In this way the effect of magnetic field on the sound attenuation could be explored.

While calculating the Prandtl number, the expressions for the thermal 
conductivity and the specific heat at constant pressure in the similar 
environment are necessary. We have recently studied 
$\kappa$~\cite{Conductivities}, so we closely follow our results in 
appendix \ref{A.T.C.}. Next we have obtained $C_p$ from the 
following thermodynamic relation,
\begin{equation}\label{cp}
C_p=\frac{\partial(\varepsilon+P)}{\partial T}
~,\end{equation}
which has been calculated from the energy density and pressure 
in the similar environment. Thus we get the 
expressions of $C_p$ for isotropic, expansion-driven anisotropic and 
$B$-driven anisotropic mediums as
\begin{eqnarray}
\nonumber C_p^{\rm iso} &=& \frac{\beta^2}{3\pi^2}\sum_i g_i \int d{\rm p}~{\rm p}^2\left({\rm p}^2+3\omega_i^2\right)f_i^{\rm iso}(1-f_i^{\rm iso}) \\ && +\frac{\beta^2}{6\pi^2}g_g\int d{\rm p}~{\rm p}^2\left({\rm p}^2+3\omega_g^2\right)f_g^{\rm iso}(1+f_g^{\rm iso}), \label{iso.cp} \\ 
\nonumber C_{p,\rm ex}^{\rm aniso} &=& C_p^{\rm iso}+\frac{\xi\beta^2}{18\pi^2}\sum_i g_i \int d{\rm p} ~ \frac{{\rm p}^4}{\omega_i^2}\left({\rm p}^2+3\omega_i^2\right)f_i^{\rm iso}(1-f_i^{\rm iso}) \\ && \nonumber-\frac{\xi\beta^3}{18\pi^2}\sum_i g_i \int d{\rm p} ~ \frac{{\rm p}^4}{\omega_i}\left({\rm p}^2+3\omega_i^2\right)f_i^{\rm iso}(1-f_i^{\rm iso})(1-2f_i^{\rm iso}) \\ && \nonumber+\frac{\xi\beta^2}{36\pi^2}g_g\int d{\rm p} ~ \frac{{\rm p}^4}{\omega_g^2}\left({\rm p}^2+3\omega_g^2\right)f_g^{\rm iso}(1+f_g^{\rm iso}) \\ && \nonumber-\frac{\xi\beta^3}{36\pi^2}g_g\int d{\rm p} ~ \frac{{\rm p}^4}{\omega_g}\left({\rm p}^2+3\omega_g^2\right)f_g^{\rm iso}(1+f_g^{\rm iso})(1+2f_g^{\rm iso}) \\ &=& \nonumber C_p^{\rm iso}+\xi\left[\frac{\beta^3}{18\pi^2}\sum_i g_i \int d{\rm p} ~ \frac{{\rm p}^4}{\omega_i}\left({\rm p}^2+3\omega_i^2\right)f_i^{\rm iso}(1-f_i^{\rm iso})\left\lbrace\frac{1}{\beta\omega_i}-1+2f_i^{\rm iso}\right\rbrace\right. \\ && \left.+\frac{\beta^3}{36\pi^2}g_g\int d{\rm p} ~ \frac{{\rm p}^4}{\omega_g}\left({\rm p}^2+3\omega_g^2\right)f_g^{\rm iso}(1+f_g^{\rm iso})\left\lbrace\frac{1}{\beta\omega_g}-1-2f_g^{\rm iso}\right\rbrace\right], \label{exaniso.cp} \\ 
\nonumber C_{p,\rm B}^{\rm aniso} &=& \frac{\beta^2}{2\pi^2}\sum_i g_i|q_iB|\int d p_3\left(p_3^2+\omega_i^2\right)f_i^{\xi=0}(1-f_i^{\xi=0}) \\ && \nonumber+\frac{\xi\beta^2}{4\pi^2}\sum_i g_i|q_iB|\int d p_3 ~ \frac{p_3^2}{\omega_i^2}\left(p_3^2+\omega_i^2\right)f_i^{\xi=0}(1-f_i^{\xi=0}) \\ && \nonumber-\frac{\xi\beta^3}{4\pi^2}\sum_i g_i|q_iB|\int d p_3 ~ \frac{p_3^2}{\omega_i}\left(p_3^2+\omega_i^2\right)f_i^{\xi=0}(1-f_i^{\xi=0})(1-2f_i^{\xi=0}) \\ && \nonumber+\frac{\beta^2}{6\pi^2}g_g\int d{\rm p}~{\rm p}^2\left({\rm p}^2+3\omega_g^2\right)f_g^{\rm iso}(1+f_g^{\rm iso}) \\ &=& \nonumber C_p^{\xi=0}+\xi\left[\frac{\beta^3}{4\pi^2}\sum_i g_i|q_iB|\int d p_3 ~ \frac{p_3^2}{\omega_i}\left(p_3^2+\omega_i^2\right)f_i^{\xi=0}(1-f_i^{\xi=0})\right. \\ && \left.\hspace{7.3 cm}\times\left\lbrace\frac{1}{\beta\omega_i}-1+2f_i^{\xi=0}\right\rbrace\right] \label{baniso.cp}
,\end{eqnarray}
respectively. 

\begin{figure}[]
\begin{center}
\includegraphics[width=7cm]{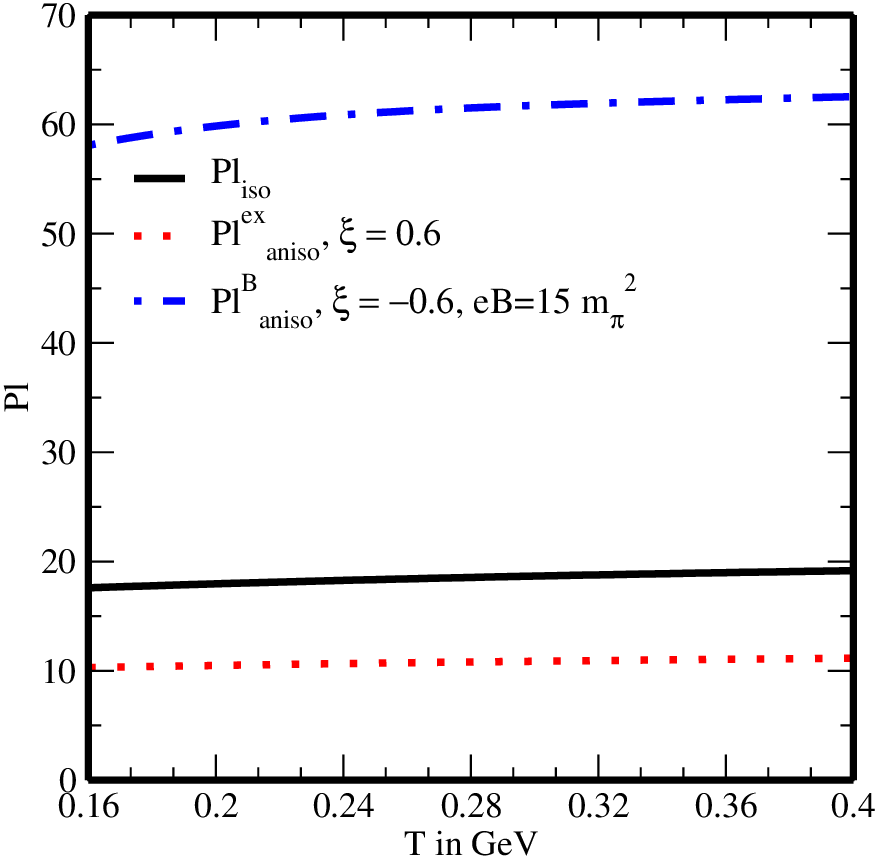}
\caption{Variation of the Prandtl number 
with temperature in the presence of momentum anisotropies 
both due to asymptotic expansion and strong 
magnetic field.}\label{pl.1}
\end{center}
\end{figure}

Finally the mass density ($\rho$) has been obtained from the product 
of the number densities of quarks and gluons with the respective 
quasiparticle masses as
\begin{equation}\label{M.D.}
\rho=2\sum_i m_in_i+m_gn_g
~.\end{equation}
The factor ``2'' represents the equal contributions from quark and 
antiquark due to $\mu_q=0$. Therefore, we get the expressions of $\rho$ 
for isotropic, expansion-driven anisotropic and $B$-driven anisotropic mediums as
\begin{eqnarray}
\rho^{\rm iso} &=& \frac{1}{\pi^2}\sum_i m_i g_i \int d{\rm p}~{\rm p}^2f_i^{\rm iso}+\frac{1}{2\pi^2}m_g g_g\int d{\rm p}~{\rm p}^2f_g^{\rm iso}, \label{iso.(M.D.)} \\ 
\nonumber\rho_{\rm ex}^{\rm aniso} &=& \rho^{\rm iso}-\frac{\xi\beta}{6\pi^2}\sum_i m_i g_i \int d{\rm p}~\frac{{\rm p}^4}{\omega_i}f_i^{\rm iso}(1-f_i^{\rm iso})-\frac{\xi\beta}{12\pi^2}m_g g_g\int d{\rm p}~\frac{{\rm p}^4}{\omega_g}f_g^{\rm iso}(1+f_g^{\rm iso}) \\ &=& \nonumber\rho^{\rm iso}-\xi\left[\frac{\beta}{6\pi^2}\sum_i m_i g_i \int d{\rm p}~\frac{{\rm p}^4}{\omega_i}f_i^{\rm iso}(1-f_i^{\rm iso})\right. \\ && \left.+\frac{\beta}{12\pi^2}m_g g_g\int d{\rm p}~\frac{{\rm p}^4}{\omega_g}f_g^{\rm iso}(1+f_g^{\rm iso})\right], \label{exaniso.(M.D.)} \\ 
\nonumber \rho_{\rm B}^{\rm aniso} &=& \frac{1}{2\pi^2}\sum_i m_i g_i|q_iB|\int d p_3 f_i^{\xi=0}-\frac{\xi\beta}{4\pi^2}\sum_i m_i g_i|q_iB|\int d p_3 ~ \frac{p_3^2}{\omega_i}f_i^{\xi=0}(1-f_i^{\xi=0}) \\ && \nonumber+\frac{1}{2\pi^2}m_g g_g\int d{\rm p}~{\rm p}^2f_g^{\rm iso} \\ &=& \rho^{\xi=0}-\frac{\xi\beta}{4\pi^2}\sum_i m_i g_i|q_iB|\int d p_3 ~ \frac{p_3^2}{\omega_i}f_i^{\xi=0}(1-f_i^{\xi=0}) \label{baniso.(M.D.)}
,\end{eqnarray}
respectively. We have therefore computed the Prandtl number as a function 
of temperature (seen in figure \ref{pl.1}) and this is found to increase 
very slowly with the temperature. It maintains higher magnitude in 
$B$-driven anisotropy than in isotropic medium and expansion-driven 
anisotropic medium as well. In all cases Prandtl number remains greater than 1, 
implying that the sound attenuation is mostly governed by the momentum diffusion.

\subsection{Reynolds number}
\begin{figure}[]
\begin{center}
\begin{tabular}{c c}
\includegraphics[width=7cm]{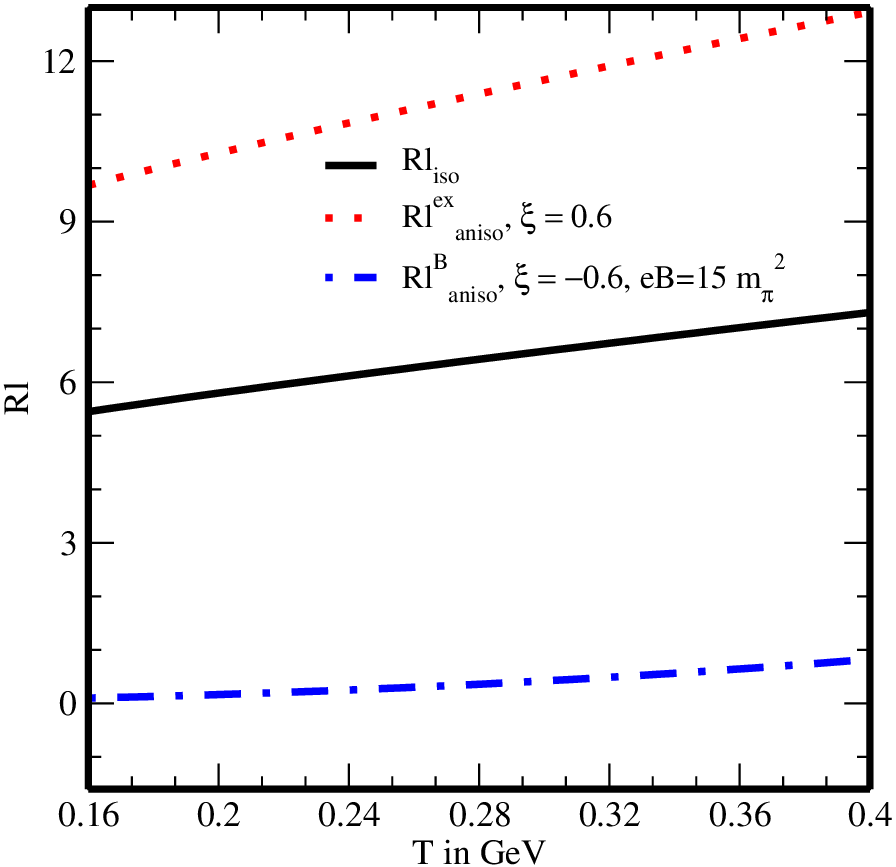}
\end{tabular}
\caption{Variation of the Reynolds number 
with temperature in the presence of momentum anisotropies 
both due to asymptotic expansion and strong 
magnetic field for $L=3$ fm.}\label{rl.1}
\end{center}
\end{figure}

The Reynolds number plays a fundamental role in 
determining the magnitude of the kinematic 
viscosity (${\eta}/{\rho}$) as compared to the 
length and velocity of the flow of a liquid and is defined by
\begin{equation}\label{Rl}
{\rm Rl}=\frac{Lv}{\eta/\rho}
~,\end{equation}
where $L$ and $v$ are the characteristic length and 
velocity of the flow, respectively. From hydrodynamic 
point of view, the Reynolds number describes the motion 
of the fluid and when the nature of the flow gets converted 
from laminar into turbulent. This conversion happens 
when ${\rm Rl}$ is much larger than 1 or kinematic viscosity 
is very small in comparison to the product of characteristic 
length and velocity ($Lv$) \cite{McInnes:NPB921'2017}. 
In (3+1)-dimensional fluid dynamical model with 
globally symmetric, peripheral initial conditions, the value 
of the Rl is estimated in the range 3-10 for initial QGP with 
minimal viscosity to entropy density ratio, {\em i.e.}, 
for $\eta/s=0.1$ \cite{Csernai:PRC85'2012}, whereas the 
holographic model reports its upper bound as approximately 20 
\cite{McInnes:NPB921'2017}. In this work, we have estimated 
the Reynolds number for (isotropic) thermal medium of 
quarks and gluons in kinetic theory approach in 
figure \ref{rl.1}, which ranges 5.5 - 7 in the temperature 
range, 160 - 400 MeV (denoted by solid line). In addition, we 
have also estimated Rl for the same but it now exhibits 
momentum anisotropies, where the expansion-driven anisotropy 
enhances the number and the $B$-driven anisotropy does the 
opposite and that too makes it less than one 
(labelled as dotted and dashed-dotted lines, respectively), 
compared to the isotropic case. 

\subsection{Relative behavior between momentum diffusion and charge diffusion}
To understand the dominance of the momentum diffusion over the charge 
diffusion, one needs to estimate the ratio of the two dimensionless
ratios: the first one is $\eta/s$ and the second one is $\sigma_{\rm el}/T$,
representing the momentum and charge diffusions, respectively. Thus, the 
ratio is given by
\begin{eqnarray}
\mathcal{\gamma}=\frac{\eta/s}{\sigma_{el}/T}
~,\end{eqnarray}
where $\sigma_{\rm el}$ is the electrical conductivity. 
Unlike gluons, only quarks carry electric charge, hence 
they only contribute to the charge transport and thus 
contribute to the electrical conductivity. On the 
other hand, both quarks and gluons participate in the 
momentum transport, thus contribute to the shear 
viscosity. Therefore, for a QGP medium, $\sigma_{\rm el}/T$ 
is always smaller than $\eta/s$, resulting the ratio, 
$\gamma$ larger than 1. This understanding is evidenced 
in ref. \cite{Puglisi:PLB751'2015}, where it is found that the large 
scattering rates due to abundance of gluons in high temperature 
QGP (compared to quarks) can damp the electrical conductivity 
and it results in the enhancement of the ratio, $\gamma$. We now 
wish to compute $\gamma$ for the hot QCD matter in the presence 
of anisotropies and also to observe the effect of strong magnetic 
field, using the kinetic theory approach. Therefore, we need to 
have the ratio, $\sigma_{\rm el}/T$ in the identical environment, 
which has been recently calculated by us \cite{Conductivities}. So, 
we closely follow our earlier calculation in appendix \ref{A.E.C.}.

In figure \ref{se.1}, we have plotted $\gamma$ 
({\em i.e.}, $(\eta/s)/(\sigma_{el}/T)$) as a function 
of temperature for isotropic medium as well as for 
expansion-driven and $B$-driven anisotropic mediums. The 
ratio $\eta/s$ is influenced by both gluon-gluon and 
quark-quark scatterings, while 
$\sigma_{el}/T$ is influenced only by the quark-quark 
scattering as only charged particles contribute to the 
electrical conductivity. Thus, the variation of $\gamma$ 
with temperature can explain the contest between gluon 
and quark contributions to the total scattering 
cross section. We have found that for an isotropic 
medium, $\gamma$ (denoted by the solid line) is maximum around $T_c$ 
($T_c=0.16$ GeV) and decreases very slowly with temperature. This is 
due to the fact that although the magnitude of $\eta/s$ is higher than 
$\sigma_{\rm el}/T$ but the latter increases relatively faster than the 
former. In the presence of expansion-driven anisotropy (denoted by dotted line), 
$\gamma$ becomes smaller than the isotropic case which is due to the 
relative decrease of $\eta/s$ than $\sigma_{\rm el}/T$ caused 
by the anisotropy. On the contrary, in the presence of strong magnetic 
field, the ratio becomes much larger than the isotropic case, which could 
be understood as follows: Although the gluon phase-space remains unaltered, 
but the quark phase-space gets reduced severely in a strong magnetic field, 
resulting an overall decrease in total entropy density. Hence $\eta/s$ 
gets enhanced by two orders of magnitude. On the other hand, the 
large increase of collisional relaxation time in strong magnetic field 
compensates the reduction in quark phase-space, resulting an increase in 
$\sigma_{\rm el}/T$ ratio, but it is now increased by one order of 
magnitude. Therefore the ratio, $\gamma$ gets increased by one order of 
magnitude. In brief, $\gamma$ remains larger than unity, so the momentum 
diffusion prevails over the charge diffusion. 

\begin{figure}[]
\begin{center}
\includegraphics[width=7cm]{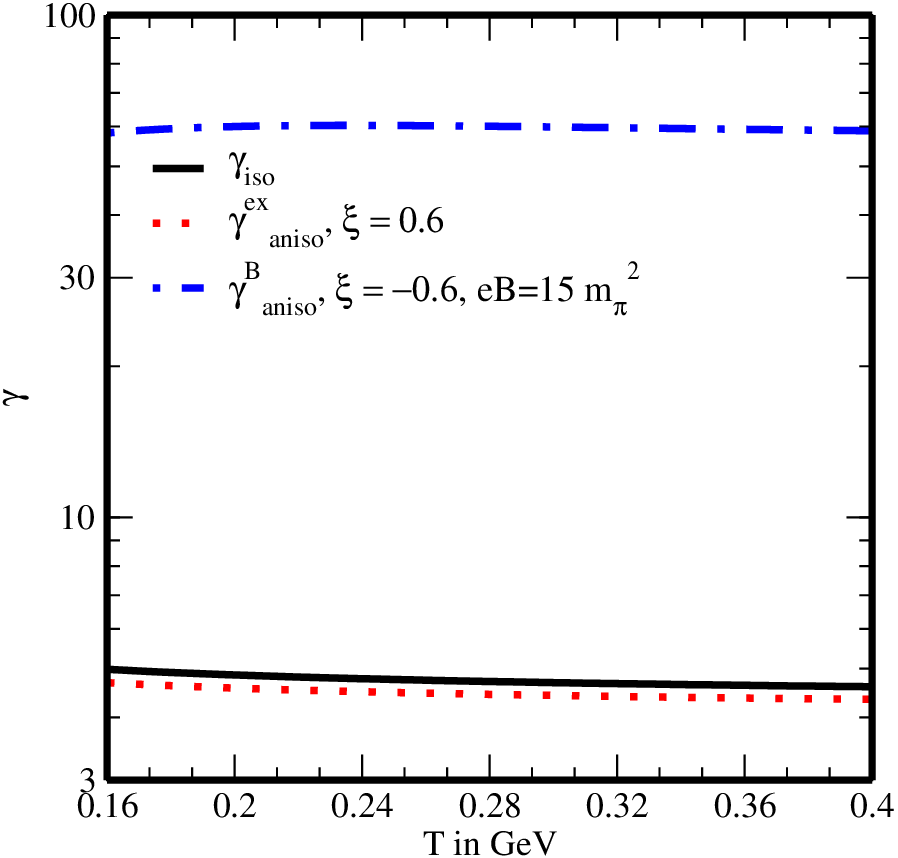}
\caption{Variation of $\gamma=(\eta/s)/(\sigma_{el}/T)$ 
with temperature in the presence of momentum anisotropies 
both due to asymptotic expansion and strong 
magnetic field.}\label{se.1}
\end{center}
\end{figure}

\section{Conclusions}
In the present work, we have first studied the momentum transports 
through the shear and bulk viscosities of a hot QCD matter and then the 
interplays among momentum, charge and heat transports are delved 
by the Prandtl number, the Reynold number and the relative behavior 
between momentum diffusion and charge diffusion. 
Most  importantly, the abovementioned studies have been 
extended to the medium with weak momentum anisotropies, which 
in turn explore the effects of strong magnetic field and asymptotic 
expansion which thought to be present at the initial stages of 
ultrarelativistic heavy ion collisions. 
We have calculated the aforesaid coefficients in the kinetic theory approach 
via the relativistic Boltzmann transport equation 
in the relaxation-time approximation and the interactions among partons are 
subsumed through the quasiparticle masses at finite temperature and strong 
magnetic field.

For that purpose, we have started with computing the shear and bulk 
viscosities in the absence and presence of expansion- and $B$-driven 
anisotropies of a thermal QCD medium. Overall observation is that the
presence of anisotropy due to strong magnetic field enhances both $\eta$ 
and $\zeta$ substantially, facilitating the transports of momentum across 
and along the layer, compared to either isotropic scenario or 
expansion-driven anisotropic scenario. Moreover, the aforesaid anisotropies 
affect $\eta$ and $\zeta$ differently with respect to the isotropic medium 
as a reference, therefore the viscosities can in principle distinguish the 
abovementioned anisotropies. Next we have computed the $\eta/s$ and
$\zeta/s$ ratios to see how the fluidity and the location of the
transition point (related to the chiral symmetry) get affected by 
the anisotropies, respectively. This enriches a competition 
between the enhancement of momentum transport and the reduction of 
phase-space (entropy density) in the presence of $B$-induced anisotropy, 
resulting the ratios, $\eta/s$ and $\zeta/s$ much greater than one, but 
unlike $\eta$ and $\zeta$, the ratios now decrease with the temperature. 
On the other hand, in the presence of expansion-driven anisotropy, both 
ratios become much smaller and specifically $\zeta/s$ vanishes around $T=0.28$ GeV. 

In the next part, we have looked into the interplay of transports between 
momentum and heat by the Prandtl number (Pl), between momentum and size of the
medium by the Reynolds number (Rl) and between momentum and charge by
the ratio, $\gamma$ in the presence of anisotropies.
The presence of strong magnetic field makes Pl much larger than its values 
in the absence of magnetic field (isotropic) as well as expansion-driven
anisotropy. Thus, in the strong magnetic field regime, the sound attenuation 
is mostly governed by the momentum diffusion. 
However, the magnetic field drops the Reynolds number to the value less than unity, 
{\em i.e.}, the kinematic viscosity dominates over the characteristic 
length and velocity of the system, which is just opposite to the effect 
caused by the expansion-driven anisotropy. Our final observation is that the 
dominance of momentum diffusion over charge diffusion is more pronounced in strong 
magnetic field than in other scenarios. However, the former one always prevails 
over the latter one. 

\section{Acknowledgment}
One of us (B. K. P.) is thankful to Council of Scientific and 
Industrial Research (Grant No. 03(1407)/17/EMR-II) for the 
financial support of this work.

\appendix
\appendixpage
\addappheadtotoc
\begin{appendices}
\renewcommand{\theequation}{A.\arabic{equation}}
\section{Thermal quark mass at finite magnetic field}\label{A.T.M.}
To compute the self-energy \eqref{Q.S.E.} at finite temperature, 
we have obtained the forms of quark and gluon propagators at 
finite temperature in the imaginary time formalism, where the 
continuous energy integral ($\int\frac{dp_0}{2\pi}$) is 
replaced by the discrete Matsubara frequency sum. Due the 
presence of strong magnetic field (along $z$-direction), the transverse 
component of momentum $k_\perp \approx0$, so, 
$e^{-k^2_\perp/|q_iB|}$ in eq. \eqref{q. propagator} 
becomes unity and the integration over the transverse component of the 
momentum gives the factor $|q_iB|$. So the quark self-energy 
\eqref{Q.S.E.} in the SMF limit takes the following form,
\begin{eqnarray}\label{Q.S.E.(1)}
\nonumber\Sigma(p_\parallel) &=& \frac{2g^2}{3\pi^2}|q_iB|T\sum_n\int dk_z\frac{\left[\left(1+\gamma^0\gamma^3\gamma^5\right)\left(\gamma^0k_0
-\gamma^3k_z\right)-2m_i\right]}{\left[k_0^2-\omega^2_k\right]\left[(p_0-k_0)^2-\omega_{pk}^2\right]} \\ &=& \frac{2g^2|q_iB|}{3\pi^2}\int dk_z\left[(\gamma^0+\gamma^3\gamma^5)L^1-(\gamma^3+\gamma^0\gamma^5)k_zL^2\right]
,\end{eqnarray}
where $\omega^2_k=k_z^2+m_i^2$, $\omega_{pk}^2=(p_z-k_z)^2$, and 
$L^1$ and $L^2$ represent two frequency sums, whose forms are given by
\be
&&L^1=T\sum_n~\frac{k_0}{\left[k_0^2-\omega_k^2\right]\left[(p_0-k_0)^2-\omega_{pk}^2\right]} ~, \\ &&L^2=T\sum_n\frac{1}{\left[k_0^2-\omega_k^2\right]\left[(p_0-k_0)^2-\omega_{pk}^2\right]}
~.\ee
After using the values of the above frequency sums, the form 
of the self-energy (\ref{Q.S.E.(1)}) turns out to be
\begin{equation}\label{Q.S.E.(2)}
\Sigma(p_\parallel)=\frac{g^2|q_iB|}{3\pi^2}\int \frac{dk_z}{\omega_k}\left[\frac{1}{e^{\beta\omega_k}-1}+\frac{1}{e^{\beta\omega_k}+1}\right]\left[\frac{\gamma^0p_0+\gamma^3p_z}{p_\parallel^2}+\frac{\gamma^0\gamma^5p_z+\gamma^3\gamma^5p_0}{p_\parallel^2}\right]
,\end{equation}
which after the integration over $k_z$, becomes
\begin{eqnarray}\label{Q.S.E.(3)}
\Sigma(p_\parallel)=\frac{g^2|q_iB|}{3\pi^2}\left[\frac{\pi T}{2m_i}-\ln(2)\right]\left[\frac{\gamma^0p_0}{p_\parallel^2}+\frac{\gamma^3p_z}{p_\parallel^2}+\frac{\gamma^0\gamma^5p_z}{p_\parallel^2}+\frac{\gamma^3\gamma^5p_0}{p_\parallel^2}\right]
.\end{eqnarray}

The covariant structure of the quark self-energy at 
finite temperature and finite magnetic field is written as
\begin{equation}\label{general q.s.e.}
\Sigma(p_\parallel)=A\gamma^\mu u_\mu+B\gamma^\mu b_\mu+C\gamma^5\gamma^\mu u_\mu+D\gamma^5\gamma^\mu b_\mu
~,\end{equation}
where $A$, $B$, $C$ and $D$ denote the form factors, and 
$u^\mu$ (1,0,0,0) and $b^\mu$  (0,0,0,-1) represent the 
preferred directions of the heat bath and the magnetic 
field, respectively. Due to the introduction of these 
vectors the Lorentz and rotational symmetries are broken. 
In LLL approximation, the form factors are obtained as
\begin{eqnarray}
&&A=\frac{1}{4}{\rm Tr}\left[\Sigma\gamma^\mu u_\mu\right]=\frac{g^2|q_iB|}{3\pi^2}\left[\frac{\pi T}{2m_i}-\ln(2)\right]\frac{p_0}{p_\parallel^2} ~, \\ 
&&B=-\frac{1}{4}{\rm Tr}\left[\Sigma\gamma^\mu b_\mu\right]=\frac{g^2|q_iB|}{3\pi^2}\left[\frac{\pi T}{2m_i}-\ln(2)\right]\frac{p_z}{p_\parallel^2} ~, \\ 
&&C=\frac{1}{4}{\rm Tr}\left[\gamma^5\Sigma\gamma^\mu u_\mu\right]=-\frac{g^2|q_iB|}{3\pi^2}\left[\frac{\pi T}{2m_i}-\ln(2)\right]\frac{p_z}{p_\parallel^2} ~, \\ 
&&D=-\frac{1}{4}{\rm Tr}\left[\gamma^5\Sigma\gamma^\mu b_\mu\right]=-\frac{g^2|q_iB|}{3\pi^2}\left[\frac{\pi T}{2m_i}-\ln(2)\right]\frac{p_0}{p_\parallel^2}
~,\end{eqnarray}
where we found that $C=-B$ and $D=-A$.

In terms of the right-handed ($P_R=(1+\gamma^5)/2$) and 
left-handed ($P_L=(1-\gamma^5)/2$) chiral projection 
operators, the quark self-energy \eqref{general q.s.e.} is written as
\begin{equation}\label{projection}
\Sigma(p_\parallel)=P_R\left[(A+C)\gamma^\mu u_\mu+(B+D)\gamma^\mu b_\mu
\right]P_L+P_L\left[(A-C)\gamma^\mu u_\mu+(B-D)\gamma^\mu b_\mu\right]P_R
~,\end{equation}
which for $C=-B$ and $D=-A$, turns out to be
\begin{equation}\label{projection1}
\Sigma(p_\parallel)=P_R\left[(A-B)\gamma^\mu u_\mu+(B-A)\gamma^\mu b_\mu
\right]P_L+P_L\left[(A+B)\gamma^\mu u_\mu+(B+A)\gamma^\mu b_\mu\right]P_R
~.\end{equation}

In the strong magnetic field regime, the effective quark propagator can 
be derived from the following self-consistent Schwinger-Dyson equation,
\be
S^{-1}(p_\parallel)=\gamma^\mu p_{\parallel\mu}-\Sigma(p_\parallel)
~,\ee
which, in terms of projection operators, is rewritten as
\be
S^{-1}(p_\parallel)=P_R\gamma^\mu X_\mu P_L+P_L\gamma^\mu Y_\mu P_R
~,\ee
where
\begin{eqnarray}
&&\gamma^\mu X_\mu=\gamma^\mu p_{\parallel\mu}-(A-B)\gamma^\mu u_\mu-(B-A)\gamma^\mu b_\mu ~, \\ 
&&\gamma^\mu Y_\mu=\gamma^\mu p_{\parallel\mu}-(A+B)\gamma^\mu u_\mu-(B+A)\gamma^\mu b_\mu
~.\end{eqnarray}
Now the effective propagator takes the following form,
\be
S(p_\parallel)=\frac{1}{2}\left[P_R\frac{\gamma^\mu Y_\mu}{Y^2/2}P_L+
P_L\frac{\gamma^\mu X_\mu}{X^2/2}P_R\right]
,\ee
where
\begin{eqnarray}
&&\frac{X^2}{2}=X_1^2=\frac{1}{2}\left[p_0-(A-B)\right]^2-\frac{1}{2}\left[p_z+(B-A)\right]^2 ~, \\ 
&&\frac{Y^2}{2}=Y_1^2=\frac{1}{2}\left[p_0-(A+B)\right]^2-\frac{1}{2}\left[p_z+(B+A)\right]^2
~.\end{eqnarray}

After taking $p_0=0, p_z\rightarrow 0$ limit of either $X_1^2$ or $Y_1^2$ 
(which are equal in this limit), we get the thermal mass (squared) at 
finite temperature and strong magnetic field as
\begin{eqnarray}
m_{iT,B}^2=X_1^2\Big{|}_{p_0=0,p_z\rightarrow 0}=Y_1^2\Big{|}_{p_0=0,p_z\rightarrow 0}=\frac{g^2|q_iB|}{3\pi^2}\left[\frac{\pi T}{2m_i}-\ln(2)\right]
.\end{eqnarray}

\renewcommand{\theequation}{B.\arabic{equation}}
\section{Form of $\Delta\tilde{T}^{ij}$ in the presence of strong magnetic field}\label{form.tensor}
In the presence of strong magnetic 
field, $\tilde{T}^{\mu\nu}$ and $\tilde{n}^\mu$ are defined as
\begin{eqnarray}
\nonumber\tilde{T}^{\mu\nu} &=& (\epsilon+P){u}^\mu{u}^\nu-Pg_\parallel^{\mu\nu}+\Delta\tilde{T}^{\mu\nu} \\ &=& \omega{u}^\mu{u}^\nu-Pg_\parallel^{\mu\nu}+\Delta\tilde{T}^{\mu\nu}, \label{EQ.1} \\ \tilde{n}^\mu &=& n{u}^\mu+\tilde{\gamma}^\mu \label{EQ.2}
,\end{eqnarray}
where $\Delta\tilde{T}^{\mu\nu}$, $\omega$, $n$ and $\tilde{\gamma}^\mu$ 
are the viscous stress tensor, the enthalpy, the particle number density 
and the dissipative correction to $\tilde{n}^\mu$, respectively in the 
presence of strong magnetic field. In addition, $g_\parallel^{\mu\nu}$ 
is defined as $g_\parallel^{\mu\nu}=(1,0,0,-1)$. Equations of motion are 
written as
\begin{eqnarray}
\frac{\partial\tilde{T}_\mu^\nu}{\partial\tilde{x}^\nu} &=& 0, \label{E.O.M.1} \\  \frac{\partial\tilde{n}^\mu}{\partial\tilde{x}^\mu} &=& 0 \label{E.O.M.2}
,\end{eqnarray}
where $\tilde{x}^\mu=(x^0,0,0,x^3)$ is redefined for the 
calculation in strong magnetic field. 

From equations \eqref{EQ.1} and \eqref{E.O.M.1}, we obtain 
\begin{eqnarray}
{u}_\mu\frac{\partial}{\partial\tilde{x}^\nu}\left(\omega{u}^\nu\right)+\omega{u}^\nu
\frac{\partial\tilde{u_\mu}}{\partial\tilde{x}^\nu}-\frac{\partial}{\partial\tilde{x}^\nu}\left(Pg_{\parallel\mu}^\nu\right)
+\frac{\partial\Delta\tilde{T}_\mu^\nu}{\partial\tilde{x}^\nu} &=& 0
.\end{eqnarray}
Now multiplying ${u}^\mu$ on both sides of the above equation and 
simplifying, we get 
\begin{eqnarray}\label{E.O.M.3}
\frac{\partial}{\partial\tilde{x}^\nu}\left(\omega{u}^\nu\right)-{u}^\nu\frac{\partial P}{\partial\tilde{x}^\nu}+{u}^\mu\frac{\partial\Delta\tilde{T}_\mu^\nu}{\partial\tilde{x}^\nu} &=& 0
.\end{eqnarray}
From equations \eqref{EQ.2} and \eqref{E.O.M.2}, we get 
\begin{eqnarray}\label{E.O.M.4}
\frac{\partial}{\partial\tilde{x}^\mu}\left(n{u}^\mu\right) &=& -\frac{\partial\tilde{\gamma}^\mu}{\partial\tilde{x}^\mu}
.\end{eqnarray}
Equation \eqref{E.O.M.4} is known as the ``equation of continuity''. 
Using the identity $\omega{u}^\nu=n{u}^\nu\frac{\omega}{n}$ in 
eq. \eqref{E.O.M.3}, we have
\begin{eqnarray}
\frac{\omega}{n}\frac{\partial}{\partial\tilde{x}^\nu}\left(n{u}^\nu\right)+n{u}^\nu\frac{\partial}{\partial\tilde{x}^\nu}\left(\frac{\omega}{n}\right)-{u}^\nu\frac{\partial P}{\partial\tilde{x}^\nu}+{u}^\mu\frac{\partial\Delta\tilde{T}_\mu^\nu}{\partial\tilde{x}^\nu} &=& 0
,\end{eqnarray}
which with the help of the equation of continuity \eqref{E.O.M.4} becomes
\begin{eqnarray}\label{simpl.1}
-\frac{\omega}{n}\frac{\partial\tilde{\gamma}^\nu}{\partial\tilde{x}^\nu}+n{u}^\nu\frac{\partial}{\partial\tilde{x}^\nu}\left(\frac{\omega}{n}\right)-{u}^\nu\frac{\partial P}{\partial\tilde{x}^\nu}+{u}^\mu\frac{\partial\Delta\tilde{T}_\mu^\nu}{\partial\tilde{x}^\nu} &=& 0
.\end{eqnarray}

With the thermodynamic relation : $d\left(\frac{\omega}{n}\right)=T
d\left(\frac{s}{n}\right)+\frac{1}{n}dP$, we have $\frac{\partial}{\partial\tilde{x}^\nu}\left(\frac{\omega}{n}\right)=T\frac{\partial}{\partial\tilde{x}^\nu}\left(\frac{s}{n}\right)+\frac{1}{n}
\frac{\partial P}{\partial\tilde{x}^\nu}$, where $s$ 
is the entropy per unit proper volume. Now, eq. \eqref{simpl.1} takes the 
following form,
\begin{eqnarray}
-\frac{\omega}{n}\frac{\partial\tilde{\gamma}^\nu}{\partial\tilde{x}^\nu}+n{u}^\nu\left[T\frac{\partial}{\partial\tilde{x}^\nu}\left(\frac{s}{n}\right)+\frac{1}{n}
\frac{\partial P}{\partial\tilde{x}^\nu}\right]-{u}^\nu\frac{\partial P}{\partial\tilde{x}^\nu}+{u}^\mu\frac{\partial\Delta\tilde{T}_\mu^\nu}{\partial\tilde{x}^\nu} &=& 0
,\end{eqnarray}
which after simplification becomes
\begin{eqnarray}\label{simpl.2}
-\frac{\omega}{n}\frac{\partial\tilde{\gamma}^\nu}{\partial\tilde{x}^\nu}+n{u}^\nu T\frac{\partial}{\partial\tilde{x}^\nu}\left(\frac{s}{n}\right)+{u}^\mu\frac{\partial\Delta\tilde{T}_\mu^\nu}{\partial\tilde{x}^\nu} &=& 0
.\end{eqnarray}
The second term in l.h.s. of the above eq. \eqref{simpl.2} can 
be written as
\begin{eqnarray}\label{deriv.1}
\nonumber n{u}^\nu T\frac{\partial}{\partial\tilde{x}^\nu}\left(\frac{s}{n}\right) &=& T\frac{\partial}{\partial\tilde{x}^\nu}\left(\frac{s}{n}n{u}^\nu\right)-T\frac{s}{n}\frac{\partial}{\partial\tilde{x}^\nu}\left(n{u}^\nu\right) \\ &=& T\frac{\partial}{\partial\tilde{x}^\nu}\left(s{u}^\nu\right)+T\frac{s}{n}\frac{\partial\tilde{\gamma}^\nu}{\partial\tilde{x}^\nu}
.\end{eqnarray}
Using eq. \eqref{deriv.1} in eq. \eqref{simpl.2}, we get
\begin{eqnarray}
-\frac{\omega}{n}\frac{\partial\tilde{\gamma}^\nu}{\partial\tilde{x}^\nu}+T\frac{\partial}{\partial\tilde{x}^\nu}\left(s{u}^\nu\right)+T\frac{s}{n}\frac{\partial\tilde{\gamma}^\nu}{\partial\tilde{x}^\nu}+{u}^\mu\frac{\partial\Delta\tilde{T}_\mu^\nu}{\partial\tilde{x}^\nu} &=& 0
.\end{eqnarray}
After rearranging the terms, the above equation tuns out to be 
\begin{eqnarray}\label{simpl.3}
\left(\frac{\omega-Ts}{n}\right)\frac{\partial\tilde{\gamma}^\nu}{\partial\tilde{x}^\nu}-T\frac{\partial}{\partial\tilde{x}^\nu}\left(s{u}^\nu\right)-{u}^\mu
\frac{\partial\Delta\tilde{T}_\mu^\nu}{\partial\tilde{x}^\nu} &=& 0
,\end{eqnarray}
where $\frac{\omega-Ts}{n}=\mu=$relativistic chemical potential. 
So, in terms of $\mu$, eq. \eqref{simpl.3} is rewritten as
\begin{eqnarray}
\frac{\partial}{\partial\tilde{x}^\nu}\left(s{u}^\nu\right)-\frac{\mu}{T}\frac{\partial\tilde{\gamma}^\nu}{\partial\tilde{x}^\nu}+\frac{{u}^\mu}{T}\frac{\partial\Delta\tilde{T}_\mu^\nu}{\partial\tilde{x}^\nu} &=& 0
,\end{eqnarray}
which can be further simplified into 
\begin{eqnarray}\label{simpl.4}
\frac{\partial}{\partial\tilde{x}^\nu}\left(s{u}^\nu\right)-\frac{\partial}{\partial\tilde{x}^\nu}\left(\frac{\mu}{T}\tilde{\gamma}^\nu\right)+\tilde{\gamma}^\nu\frac{\partial}{\partial\tilde{x}^\nu}\left(\frac{\mu}{T}\right)+\frac{1}{T}\frac{\partial}{\partial\tilde{x}^\nu}\left({u}^\mu\Delta\tilde{T}_\mu^\nu\right)
-\frac{\Delta\tilde{T}_\mu^\nu}{T}\frac{\partial{u}^\mu}{\partial\tilde{x}^\nu} = 0
.\end{eqnarray}
Using ${u}^\mu\Delta\tilde{T}_\mu^\nu=0$ in the above 
eq. \eqref{simpl.4}, we get
\begin{eqnarray}
\frac{\partial}{\partial\tilde{x}^\nu}\left(s{u}^\nu\right)-\frac{\partial}{\partial\tilde{x}^\nu}\left(\frac{\mu}{T}\tilde{\gamma}^\nu\right)+\tilde{\gamma}^\nu\frac{\partial}{\partial\tilde{x}^\nu}\left(\frac{\mu}{T}\right)-\frac{\Delta\tilde{T}_\mu^\nu}{T}\frac{\partial{u}^\mu}{\partial\tilde{x}^\nu} &=& 0
,\end{eqnarray}
which after simplification becomes 
\begin{eqnarray}\label{simpl.5}
\frac{\partial}{\partial\tilde{x}^\nu}\left(s{u}^\nu-\frac{\mu}{T}\tilde{\gamma}^\nu\right) &=& -\tilde{\gamma}^\nu\frac{\partial}{\partial\tilde{x}^\nu}\left(\frac{\mu}{T}\right)+\frac{\Delta\tilde{T}_\mu^\nu}{T}\frac{\partial{u}^\mu}{\partial\tilde{x}^\nu}
.\end{eqnarray}
In eq. \eqref{simpl.5}, $s{u}^\nu-\frac{\mu}{T}\tilde{\gamma}^\nu=\tilde{s}^\nu=$entropy flux density 4-vector 
in the presence of strong magnetic field. So, in terms 
of $\tilde{s}^\nu$, eq. \eqref{simpl.5} is rewritten as
\begin{eqnarray}\label{simpl.6}
\frac{\partial\tilde{s}^\nu}{\partial\tilde{x}^\nu} &=& -\tilde{\gamma}^\nu\frac{\partial}{\partial\tilde{x}^\nu}\left(\frac{\mu}{T}\right)+\frac{\Delta\tilde{T}_\mu^\nu}{T}\frac{\partial{u}^\mu}{\partial\tilde{x}^\nu}
.\end{eqnarray}
Here $\frac{\partial\tilde{s}^\nu}{\partial\tilde{x}^\nu}$ is the 
4-divergence of the entropy flux density in a strong magnetic 
field. According to the law of increase of entropy, the r.h.s. of 
eq. \eqref{simpl.6} must be positive. Thus, a most general form 
of $\Delta\tilde{T}^{\mu\nu}$ that satisfies 
$\Delta\tilde{T}^{\mu\nu}{u}_\nu=0$ and the law of increase 
of entropy is written as
\begin{eqnarray}\label{FORM}
\Delta\tilde{T}^{\mu\nu}=-\eta^B\left(\frac{\partial{u}^\mu}{\partial\tilde{x}_\nu}+\frac{\partial{u}^\nu}{\partial\tilde{x}_\mu}-{u}^\nu{u}_\lambda\frac{\partial{u}^\mu}{\partial\tilde{x}_\lambda}-{u}^\mu{u}_\lambda\frac{\partial{u}^\nu}{\partial\tilde{x}_\lambda}-\frac{2}{3}\Delta_\parallel^{\mu\nu}\frac{\partial{u}^\lambda}{\partial\tilde{x}^\lambda}\right)
-\zeta^B\Delta_\parallel^{\mu\nu}\frac{\partial{u}^\lambda}{\partial\tilde{x}^\lambda}
,\end{eqnarray}
where $\Delta_\parallel^{\mu\nu}=g_\parallel^{\mu\nu}-{u}^\mu{u}^\nu$, $\eta^B$ and $\zeta^B$ are the shear viscosity and the bulk viscosity, 
respectively in the presence of strong magnetic field. In the 
local rest frame, the spatial component of velocity is zero, but 
its spatial derivative remains finite. Therefore, the spatial 
component of eq. \eqref{FORM} is written as
\begin{eqnarray}\label{S.FORM}
\nonumber\Delta\tilde{T}^{ij} &=& -\eta^B\left(\frac{\partial{u}^i}{\partial\tilde{x}_j}+\frac{\partial{u}^j}{\partial\tilde{x}_i}-\frac{2}{3}\delta^{ij}\frac{\partial{u}^l}{\partial\tilde{x}^l}\right)
-\zeta^B\delta^{ij}\frac{\partial{u}^l}{\partial\tilde{x}^l} \\ &=& \nonumber -\eta^B\left(\partial^i{u}^j+\partial^j{u}^i-\frac{2}{3}\delta^{ij}\partial_l{u}^l\right)
-\zeta^B\delta^{ij}\partial_l{u}^l \\ &=& -\eta^B{W}^{ij}-\zeta^B\delta^{ij}\partial_l{u}^l
.\end{eqnarray}

\renewcommand{\theequation}{C.\arabic{equation}}
\section{Energy density and pressure}\label{A.(e,p,s)}
The thermodynamic quantities, {\em such as}, the energy density 
($\varepsilon$) and the pressure ($P$) can be obtained from the 
energy momentum tensor ($T^{\mu\nu}$). In the absence of 
magnetic field, we have
\begin{eqnarray}
\varepsilon &=& u_\mu T^{\mu\nu}u_\nu
, \\
P &=& -\frac{1}{3}\left(g_{\mu\nu}-u_\mu u_\nu\right)T^{\mu\nu}
,\end{eqnarray}
whereas in the presence of strong magnetic field, the definitions 
of the energy density and the pressure get modified as
\begin{eqnarray}
\varepsilon &=& u_\mu\tilde{T}^{\mu\nu}u_\nu
, \\
P &=& -\left(g^\parallel_{\mu\nu}-u_\mu u_\nu\right)\tilde{T}^{\mu\nu}
.\end{eqnarray}

Expressions of energy density for isotropic, expansion-driven anisotropic and 
$B$-driven anisotropic mediums are calculated as
\begin{eqnarray}
\varepsilon^{\rm iso} &=& \frac{1}{\pi^2}\sum_i g_i \int d{\rm p}~{\rm p}^2\omega_if_i^{\rm iso}+\frac{1}{2\pi^2}g_g\int d{\rm p}~{\rm p}^2\omega_gf_g^{\rm iso}
, \label{iso.(Er.D.)} \\ 
\nonumber \varepsilon_{\rm ex}^{\rm aniso} &=& \varepsilon^{\rm iso}-\frac{\xi\beta}{6\pi^2}\sum_i g_i \int d{\rm p}~{{\rm p}^4}f_i^{\rm iso}(1-f_i^{\rm iso}) \\ && \nonumber-\frac{\xi\beta}{12\pi^2}g_g\int d{\rm p}~{{\rm p}^4}f_g^{\rm iso}(1+f_g^{\rm iso}) \\ &=& \nonumber \varepsilon^{\rm iso}-\xi\left[\frac{\beta}{6\pi^2}\sum_i g_i \int d{\rm p}~{{\rm p}^4}f_i^{\rm iso}(1-f_i^{\rm iso})\right. \\ && \left.+\frac{\beta}{12\pi^2}g_g\int d{\rm p}~{{\rm p}^4}f_g^{\rm iso}(1+f_g^{\rm iso})\right]
, \label{exaniso.(Er.D.)} \\ 
\nonumber \varepsilon_{\rm B}^{\rm aniso} &=& \frac{1}{2\pi^2}\sum_i g_i|q_iB|\int d p_3\omega_if_i^{\xi=0} \\ && \nonumber-\frac{\xi\beta}{4\pi^2}\sum_i g_i|q_iB|\int d p_3 ~ {p_3^2}f_i^{\xi=0}(1-f_i^{\xi=0}) \\ && \nonumber+\frac{1}{2\pi^2}g_g\int d{\rm p}~{\rm p}^2\omega_gf_g^{\rm iso} \\ &=& \varepsilon^{\xi=0}-\frac{\xi\beta}{4\pi^2}\sum_i g_i|q_iB|\int d p_3 ~ {p_3^2}f_i^{\xi=0}(1-f_i^{\xi=0}) \label{baniso.(Er.D.)}
,\end{eqnarray}
respectively.

Expressions of pressure for isotropic, expansion-driven anisotropic and 
$B$-driven anisotropic mediums are calculated as
\begin{eqnarray}
P^{\rm iso} &=& \frac{1}{3\pi^2}\sum_i g_i \int d{\rm p}~\frac{{\rm p}^4}{\omega_i}f_i^{\rm iso}+\frac{1}{6\pi^2}g_g\int d{\rm p}~\frac{{\rm p}^4}{\omega_g}f_g^{\rm iso}
, \label{iso.(P.)} \\ 
\nonumber P_{\rm ex}^{\rm aniso} &=& P^{\rm iso}-\frac{\xi\beta}{18\pi^2}\sum_i g_i \int d{\rm p}~\frac{{\rm p}^6}{\omega_i^2}f_i^{\rm iso}(1-f_i^{\rm iso}) \\ && \nonumber-\frac{\xi\beta}{36\pi^2}g_g\int d{\rm p}~\frac{{\rm p}^6}{\omega_g^2}f_g^{\rm iso}(1+f_g^{\rm iso}) \\ &=& \nonumber P^{\rm iso}-\xi\left[\frac{\beta}{18\pi^2}\sum_i g_i \int d{\rm p}~\frac{{\rm p}^6}{\omega_i^2}f_i^{\rm iso}(1-f_i^{\rm iso})\right. \\ && \left.+\frac{\beta}{36\pi^2}g_g\int d{\rm p}~\frac{{\rm p}^6}{\omega_g^2}f_g^{\rm iso}(1+f_g^{\rm iso})\right]
, \label{exaniso.(P.)} \\ 
\nonumber P_{\rm B}^{\rm aniso} &=& \frac{1}{2\pi^2}\sum_i g_i|q_iB|\int d p_3\frac{p_3^2}{\omega_i}f_i^{\xi=0} \\ && \nonumber-\frac{\xi\beta}{4\pi^2}\sum_i g_i|q_iB|\int d p_3 ~ \frac{p_3^4}{\omega_i^2}f_i^{\xi=0}(1-f_i^{\xi=0}) \\ && \nonumber+\frac{1}{6\pi^2}g_g\int d{\rm p}~\frac{{\rm p}^4}{\omega_g}f_g^{\rm iso} \\ &=& P^{\xi=0}-\frac{\xi\beta}{4\pi^2}\sum_i g_i|q_iB|\int d p_3 ~ \frac{p_3^4}{\omega_i^2}f_i^{\xi=0}(1-f_i^{\xi=0}) \label{baniso.(P.)}
,\end{eqnarray}
respectively.

\renewcommand{\theequation}{D.\arabic{equation}}
\section{Thermal conductivity}\label{A.T.C.}
For isotropic medium, thermal conductivity is given by
\begin{eqnarray}\label{iso.}
\kappa^{\rm iso} = \frac{\beta^2}{3\pi^2}\sum_ig_i\int d{\rm p} \frac{{\rm p}^4}{\omega_i^2}(\omega_i-h_i)^2 ~ \tau_i ~ f_i^{\rm iso}(1-f_i^{\rm iso})
.\end{eqnarray}
For expansion-driven anisotropic medium, thermal conductivity is given by
\begin{eqnarray}\label{ex.}
\nonumber\kappa_{\rm ex}^{\rm aniso} &=& \kappa^{\rm iso}+\xi\left[\frac{\beta^2}{18\pi^2}\sum_ig_i\int d{\rm p} ~ \frac{{\rm p}^6}{\omega_i^4}(\omega_i^2-h_i^2) ~ \tau_i ~ f_i^{\rm iso}(1-f_i^{\rm iso})\right. \\ && \left.-\frac{\beta^3}{18\pi^2}\sum_ig_i\int d{\rm p} ~ \frac{{\rm p}^6}{\omega_i^3}(\omega_i-h_i)^2 ~ \tau_i ~ f_i^{\rm iso}(1-2f_i^{\rm iso})(1-f_i^{\rm iso})\right]
.\end{eqnarray}
For $B$-driven anisotropic medium, thermal conductivity is given by
\begin{eqnarray}
\kappa_{\rm B}^{\rm aniso} &=& \nonumber\frac{\beta^2}{2\pi^2}\sum_ig_i|q_iB|\int dp_3 ~ \frac{p_3^2}{\omega_i^2}(\omega_i-h_i^B)^2 ~ \tau_i^B ~ f_i^{\xi=0}(1-f_i^{\xi=0}) \\ && \nonumber+\frac{\xi\beta^2}{4\pi^2}\sum_ig_i|q_iB|\int dp_3 ~ \frac{p_3^4}{\omega_i^4}(\omega_i^2-{h_i^B}^2) ~ \tau_i^B ~ f_i^{\xi=0}(1-f_i^{\xi=0}) \\ && \nonumber -\frac{\xi\beta^3}{4\pi^2}\sum_ig_i|q_iB|\int dp_3 ~ \frac{p_3^4}{\omega_i^3}(\omega_i-h_i^B)^2 ~ \tau_i^B ~ f_i^{\xi=0}(1-2f_i^{\xi=0}) \\ && \hspace{7.8 cm}\times(1-f_i^{\xi=0})
.\end{eqnarray}
This can be decomposed into $\xi=0$ and $\xi\neq 0$ parts as
\begin{eqnarray}\label{eb}
\nonumber\kappa_{\rm B}^{\rm aniso} &=& \kappa^{\xi=0}+\kappa^{\xi\ne 0} \\ &=& \nonumber\kappa^{\xi=0}+\xi\left[\frac{\beta^2}{4\pi^2}\sum_ig_i|q_iB|\int dp_3 ~ \frac{p_3^4}{\omega_i^4}(\omega_i^2-{h_i^B}^2) ~ \tau_i^B ~ f_i^{\xi=0}(1-f_i^{\xi=0})\right. \\ && \left.\nonumber -\frac{\beta^3}{4\pi^2}\sum_ig_i|q_iB|\int dp_3 ~ \frac{p_3^4}{\omega_i^3}(\omega_i-h_i^B)^2 ~ \tau_i^B ~ f_i^{\xi=0}(1-2f_i^{\xi=0})\right. \\ && \left.\hspace{7.8 cm}\times(1-f_i^{\xi=0})\right]
.\end{eqnarray}

\renewcommand{\theequation}{E.\arabic{equation}}
\section{Electrical conductivity}\label{A.E.C.}
For isotropic medium, electrical conductivity is given by
\begin{eqnarray}\label{I.E.C.}
\sigma_{\rm el}^{\rm iso}=\frac{2\beta}{3\pi^2}\sum_i g_i q_i^2\int d{\rm p}~\frac{{\rm p}^4}{\omega_i^2} ~ \tau_i ~ f_i^{\rm iso}(1-f_i^{\rm iso})
.\end{eqnarray}
For expansion-driven anisotropic medium, electrical conductivity is given by
\begin{eqnarray}\label{A.E.C.(1)}
\nonumber\sigma_{\rm el,ex}^{\rm aniso} &=& \sigma_{\rm el}^{\rm iso} - \xi\left[\frac{\beta^2}{9\pi^2}\sum_i g_i q_i^2\int d{\rm p}\frac{{\rm p}^6}{\omega_i^3}~ \tau_i ~ f_i^{\rm iso}(1-f_i^{\rm iso})\left\lbrace 1-2f_i^{\rm iso}+\frac{1}{\beta\omega_i} \right\rbrace\right. \\ && \left.-\frac{\beta}{9\pi^2}\sum_i g_i q_i^2\int d{\rm p}\frac{{\rm p}^4}{\omega_i^2}~ \tau_i ~ f_i^{\rm iso}(1-f_i^{\rm iso})\right]
.\end{eqnarray}
For $B$-driven anisotropic medium, electrical conductivity is given by
\begin{eqnarray}\label{A.E.C.(eB)}
\nonumber\sigma_{\rm el,B}^{\rm aniso} &=& \frac{\beta}{\pi^2}\sum_i g_i q_i^2~|q_iB|\int dp_3~\frac{p_3^2}{\omega_i^2} ~ \tau_i^B ~ f_i^{\xi=0}(1-f_i^{\xi=0}) \\ && - \nonumber\frac{\xi\beta^2}{2\pi^2}\sum_i g_i q_i^2~|q_iB|\int d{p_3}~\frac{p_3^4}{\omega_i^3}~ \tau_i^B ~ f_i^{\xi=0}(1-f_i^{\xi=0})\left\lbrace 1-2f_i^{\xi=0}+\frac{1}{\beta\omega_i} \right\rbrace \\ && +\frac{\xi\beta}{2\pi^2}\sum_i g_i q_i^2~|q_iB|\int d{p_3}~\frac{p_3^2}{\omega_i^2}~ \tau_i^B ~ f_i^{\xi=0}(1-f_i^{\xi=0})
.\end{eqnarray}
This can be decomposed into $\xi=0$ and $\xi\neq 0$ parts as
\begin{eqnarray}\label{A.E.C.(1eB)}
\nonumber\sigma_{\rm el,B}^{\rm aniso} &=& \sigma_{\rm el}^{\xi=0}+\sigma_{\rm el}^{\xi\ne 0} \\ &=& \nonumber\sigma_{\rm el}^{\xi=0} - \xi\left[\frac{\beta^2}{2\pi^2}\sum_i g_i q_i^2~|q_iB|\int d{p_3}~\frac{p_3^4}{\omega_i^3}~ \tau_i^B ~ f_i^{\xi=0}(1-f_i^{\xi=0})\left\lbrace 1-2f_i^{\xi=0}+\frac{1}{\beta\omega_i} \right\rbrace\right. \\ && \left.-\frac{\beta}{2\pi^2}\sum_i g_i q_i^2~|q_iB|\int d{p_3}~\frac{p_3^2}{\omega_i^2}~ \tau_i^B ~ f_i^{\xi=0}(1-f_i^{\xi=0})\right]
.\end{eqnarray}

\end{appendices}

\end{document}